\documentclass[runningheads]{svmult}

\usepackage{makeidx}   
\usepackage{graphicx}  
\usepackage{subeqnar}  
\usepackage{multicol}  
\usepackage{physprbb}  
\makeindex             

\usepackage{amsfonts}   
\usepackage{natbib}
\usepackage{amssymb}
\newcommand{\apjl} {ApJ}
\newcommand{\apj} {ApJ}
\newcommand{\mnras} {MNRAS}
\newcommand{\aap} {A\&A}

\newcommand{\pasj} {PASJ}
\newcommand{\solphys}  {Solar Physics}

\newcommand{\apss} {Ap\&SS}
\def\aplett{Astrophys. Lett.} %

\newcommand{\ds}{\displaystyle}

\newcommand{\derivp} [2] {\frac {\partial #1 } {\partial #2} }
\newcommand{\deriv} [2] {\frac {d #1 } {d #2} }
\newcommand{\eqn} [1] {
\begin{equation} 
#1 
\end{equation}}

\newcommand{\eqna} [1] {
\begin{eqnarray} 
#1 
\end{eqnarray}}
\newcommand{\eq} [1] {{Eq.~(\ref{#1})}}
\newcommand{\eqs} [1] {{Eqs.~(\ref{#1})}}

\def\acenA{$\alpha$~Cen~A}
\def\nat{Nature} 
\begin{document}

\title*{Stochastic excitation of acoustic modes in stars}

\toctitle{Stochastic excitation of acoustic modes in stars}

\titlerunning{Stochastic excitation of acoustic modes in stars}

\author{R. Samadi}

\institute{Observatoire de Paris, LESIA, CNRS UMR 8109, 92195, Meudon, France}


\maketitle

\begin{abstract} 
\noindent For more than ten years,  solar-like
oscillations have  been detected and frequencies measured for  a growing number of  stars with various characteristics (e.g.
different evolutionary stages, effective temperatures, gravities,
metal abundances ...).

Excitation of such oscillations is  attributed to turbulent
convection and takes place in the uppermost part of the convective envelope.
Since the pioneering work of Goldreich \& Keely (1977), more sophisticated
theoretical models of stochastic excitation were developed, which
differ from each other both by the way turbulent convection is modelled and
by the assumed sources of excitation.
We review here these different models and their underlying
approximations and assumptions. 

We emphasize how the computed mode excitation rates crucially
depend on the way turbulent convection is described but also on the
stratification and  the metal abundance of the upper layers of the star.
In turn we will show how the seismic measurements collected so far 
allow us to infer  properties of turbulent convection in stars.
\end{abstract}

\section{Introduction}

Solar $p$-modes are known to have finite lifetimes (a few days) and very
low amplitudes (a few cm/s in velocity and a few ppm in intensity). 
Their finite lifetimes result from several complex damping
processes that are so far not clearly understood. Their excitation is
attributed to turbulent convection and takes place in the upper-most
part of the Sun, which is the place
of vigorous and turbulent motions. Since the pioneering work of
\citet{Lighthill52}, we know that a turbulent medium generates
incoherent acoustic pressure fluctuations (also called acoustic ``noise''). A very
small fraction of the associated kinetic energy goes into 
to the normal
modes of the solar cavity. This small amount of energy  then is
responsible for the small observed amplitudes of the solar acoustic
modes ($p$ modes). 

In the last decade, solar-like oscillations have been detected in
numerous stars, in different evolutionary stages and with different
metallicity \citep[see recent review by][]{Bedding07}. As in the Sun, these
oscillations have rather small amplitudes and have finite
lifetimes. The excitation of such solar-like oscillations is attributed to turbulent convection and takes place in the outer
layers of stars having a convective envelope.

Measuring mode amplitudes and the mode lifetimes permits us to infer
 $\cal P$, the energy  supplied per unit time into the acoustic
modes. Deriving  $\cal P$ puts constraints on the theoretical models of mode excitation
by turbulent convection \citep{Libbrecht88}. However, as pointed-out
by \citet{Baudin05}, even for the Sun, inferring  $\cal P$ from the
seismic data is not a trivial task. For stellar seismic data,
this is even more difficult \citep{Samadi08}. We discuss here the
problems we face in deriving reliable seismic constraints on $\cal
P$.

A first attempt to explain the observed solar five minute
oscillations was carried out by \citet{Unno62}. They have considered
monopole \footnote{A monopole term is associated with a
  fluctuation of density} and dipole\footnote{A dipole term is associated
  with a fluctuation of a force} source terms that arise from an isothermal
stratified atmosphere. \cite{Stein67} has generalised
\citet{Lighthill52}'s approach to a stratified atmosphere. He
found that monopole  source terms have a negligible
contribution to the noise generation compared to
the quadrupole source term\footnote{A quadrupole term  is associated with a shear}.  Among the quadrupole source terms, the Reynolds
stress was expected to be the major source of acoustic \emph{wave} generation.
It was only at the beginning of the 1970's that solar five minutes
oscillations have been clearly identified as global resonant modes
\citep{Ulrich70,Leibacher71,Deubner75}.  
A few years later, \citet[][GK hereafter]{GK77} have proposed the
first theoretical model of stochastic excitation of acoustic \emph{modes} by
the Reynolds stress.
Since this pioneering work, different improved  models have
been developed 
\citep{Dolginov84,Balmforth92c,GMK94,Samadi00I,Chaplin05,Samadi02II,Kevin06b,Kevin08}.     
These  approaches differ from each other either in the way turbulent
convection is described or by the excitation process.  

In the present paper, we briefly review  the different main
formulations and discuss the main assumptions and
approximations on which these models are based.  
As shown by \citet{Samadi02II}, the energy supplied per time unit to the modes
by turbulent convection crucially depends on the way eddies 
are temporally correlated. 
A realistic modeling of  the eddy time-correlation at various
scale lengths then is  an important issue, which is discussed in
detail here. We will also highlight how the mean structure and the
chemical composition of the upper convective envelope influence the
mode driving.   
Finally, we will summarize how the seismic measurements obtained so
far from the ground allow us to distinguish
between different dynamical descriptions of turbulent convection.

\section{Mode energy}

We will show below how the energy of a solar-like oscillation is
related to the driving and damping process.
The mode  total energy (potential plus kinetic) is by definition the quantity:
\eqna{
E_{\rm osc} (t) & =  &  \int d^3 x \, \rho_0 \,    \vec v_{\rm osc}
^2 (\vec r,t) 
\label{E_osc}
}
where $\vec v_{\rm osc}$ is the mode velocity at the position $\vec x$,
and $\rho_0$ the mean density.

Mode damping occurs over a
time-scale much longer than that associated with the driving.
Accordingly, damping and driving can be completely decoupled in time.
 Furthermore, we assume a constant and linear damping such that
\eqn{
  \deriv{\, \vec v_{\rm osc} (t) }  {t}  = - \eta \,\vec v_{\rm osc} (t)
\label{damping}
}
where $\eta$ is the (constant) damping rate.  The time derivative in \eq{damping} is
performed over a time scale much larger than  the characteristic time
over which the driving occurs.  

Let ${\cal P}$ be the amount of energy injected per unit time into a
mode by an arbitrary source of driving (which acts over a 
time scale much shorter than $1/\eta$).
According to \eqs{E_osc} and (\ref{damping}), the variation of $E_{\rm osc}$ with time is given by:
\eqna{
\deriv{E_{\rm osc}}{t} (t) & = & {\cal P} - 2\ \eta \, E_{\rm osc} (t)
\;.
\label{balance}
}
Solar-like oscillations are known to be stable modes. As a consequence, their
energy cannot growth on a time scale much longer than the time scales
associated with the damping  and driving process. 
Accordingly, averaging \eq{balance}  over a long  time scale gives:
\eqna{
\overline{\deriv{E_{\rm osc}}{t} (t) } = 0 \;,
\label{balance_2}
}
where $\overline{()}$ refer to a time average.
From Eqs.~(\ref{balance}) and (\ref{balance_2}), we immediately derive:
\eqna{
 \overline{ E}_{\rm osc}   & =  & \overline{ {\cal P} } \, \over
 { 2 \eta } \; .
\label{balance_3}
}

We then clearly see with \eq{balance_3} that a stable mode has its 
 energy (and thus its amplitude) controlled by the balance
between the driving ($\cal P$) and the damping ($\eta$). 
Then, the major difficulties are to model the processes that are
at the origin of the driving and the damping.  
For ease of  notation, we will  drop from now on the symbol
$\overline{()}$ from $E_{\rm osc}$ and ${\cal P}$.

\section{Seismic constraints}

As we shall see later, 
the mode displacement, $\delta \vec{r}_{\rm osc}$, can be 
 written   in terms of the adiabatic eigen-displacement $\vec \xi$,  
 and an instantaneous
amplitude   $A(t)$:
\eqn{
 \delta \vec r_{\rm osc}  \equiv {1 \over 2} \, \left ( A(t) \,  \vec {\xi} (\vec r) \,
 e^{-i \omega_{\rm osc} t } +cc \right )
\label{delta_osc_2}
}
where cc means complex conjugate, $\omega_{\rm osc}$ is the mode eigenfrequency, and $A(t)$ is the instantaneous
amplitude resulting from both the driving and the damping. Note that, since the normalisation of $\xi$
is arbitrary,  the actual \emph{intrinsic} mode amplitude is fixed by the
term $A(t)$, which remains to be determined.
The  mode velocity, $\vec v_{\rm osc}$ , is then given by:
\eqn{
\vec v_{\rm osc}\, (\vec r,t) = \deriv{\delta \vec{r}_{\rm osc}}{t}  =    {1\over 2} 
(  -i \omega_{\rm osc} \, A(t) \,  \vec {\xi} (\vec r) \,  e^{-i \omega_{\rm osc} t } +cc)
\label{At}
}
where cc means complex conjugate. Note that we have neglected in
\eq{At} the time derivative of $A$.  This is justified since the mode
period ($2\pi/\omega_{\rm osc}$)  is in general much shorter than the mode
lifetime ($\sim 1/\eta$)

From Eqs.~(\ref{At}) and (\ref{E_osc}), we derive the expression for
the mean mode energy:
\eqna{
{E}_{\rm osc}  & =  &  \int d^3 x \,\rho_0 \, 
\overline { \vec v_{\rm osc} ^2    } =  {1 \over 2}
\overline{ \mid A \mid ^2 }  \, I  \, {\omega_{\rm osc}}^2 \;, 
\label{E_osc_2}
}
where
\eqn{
I \equiv   \int_0^{M} d^3 x \, \rho_0 \, \vec \xi^* \, . \, \vec \xi \label{inertia}
}
is the mode inertia. For the sake of simplicity, we will from now on only
consider radial modes.
According to \eq{At}, the mean-square surface velocity associated with a \emph{radial} mode measured at  the radius $r_h$,
 is then given by the relation 
\eqn{
\vec{v}_s^2 (r_h) =  \, \frac{1}{2}
\overline{ \mid A \mid ^2  }  \, \omega_{\rm osc}^2 \, \mid \xi_{\rm r}
(r_h) \mid^2   
\label{v_s_sq}
}  
where $\xi_{\rm r}$ is the radial component of the mode eigenfunction.
It is convenient and common to define the mode mass as the quantity:
\eqn{
{\cal M} (r_h)  \equiv  { I \over {   \mid \xi_{\rm r}  (r_h) \mid^2 }}  \label{MM}
}
where $r_h$  is the radius  in the atmosphere where the
mode is measured in velocity. 
According to \eqs{E_osc_2}, (\ref{v_s_sq}), and (\ref{MM}), we derive
the following relation:
\eqn{
{E}_{\rm osc} = {\cal M}  \, \vec{v}_s^2 
\label{E_osc_3}
}
It should be noticed, that although ${\cal M}$ and $v_s$ depend on the
choice for the radius $r_h$, ${E}_{\rm osc}$ is by definition
intrinsic to the mode (see \eq{E_osc}) and hence is independent of 
$r_h$.

Using  Eqs.~(\ref{balance_3}), and (\ref{E_osc_3}), we
finally derive:
\eqna{
\vec{v}_s^2 (r_h , \omega_{\rm osc})   & = & \mathcal P \over{ 2 \, \pi \,
    \mathcal M \, \Gamma  } 
\label{v_s}
}
where $\Gamma=\eta/\pi$ is the mode linewidth, and $\eta$ the mode damping rate.
From \eq{v_s}, one again sees that the mode surface velocity is the
result of the balance between excitation (${\cal P}$) and the damping
($\eta=\Gamma\, \pi$). However, it also depends on the mode mass
(${\cal M}$): For a given driving (${\cal P}$) and damping ($\Gamma$),
the larger the mode mass (or the mode inertia), the smaller the mode
velocity. 

When the frequency resolution and the signal-to-noise are high enough,
it is possible to resolve the mode profile and then to measure \emph{both}
$\Gamma$ and the mode height $H$ in the power spectral density
(generally given in m$^2$/Hz). In that case $v_s$ is given by the relation \citep[see e.g.][]{Baudin05}:
\eqna{
v_s^2 (r_h , \omega_{\rm osc})  & = \pi \,C_{\rm obs} \, H \, \Gamma 
\label{v_s_2} 
}
where the constant $C_{\rm obs}$ takes the observational technique and
geometrical effects into account \citep[see][]{Baudin05}.
From \eq{v_s} and (\ref{v_s_2}), one can then infer from the observations
the mode excitation rates (${\cal P}$) as:
\eqn{
{\cal P}   (\omega) = 2  \pi \, {\cal M} \, \Gamma \, v_s^2 =  2
\pi^2 \, {\cal M} \, C_{\rm obs} \,H \, \Gamma ^2  \; .
\label{pow_obs}
}
Provided that we can measure $\Gamma$ and $H$,
it is then possible to constraint ${\cal P}$. However, we point out that
the derivation of ${\cal P}$ from the observations is also based on models
since  ${\cal M}$ is required. 
Furthermore, there is a strong anti-correlation between $H$ and 
$\Gamma$ \citep[see e.g.][]{Chaplin98,Chaplin08} that can introduce
important bias. 
This anti-correlation vanishes when considering the squared mode
amplitude, $v_s^2$, since $v_s^2 \propto  H \, \Gamma$ (see \eq{v_s_2}).  However,
${\cal P}$ still depends on $\Gamma$, which is 
strongly anti-correlated with $H$. 

As an alternative to comparing theoretical results and observational
data, \cite{Chaplin05} proposed to derive $H$ from the theoretical
excitation rates, ${\cal P}$, and the observed mode line width,
$\Gamma$,  according to the relation:
\eqn{
 H = { {\cal P } \over {2 \pi^2 \, {\cal M} \, C_{\rm obs} \,\Gamma ^2
   } } 
\label{H}
}
However, as  pointed-out by \cite{Kevin06b},  $H$ strongly depends on the
observation technique.
The quantity $ C_{\rm obs} \,H $, is less dependent on the
observational data  but still depends on the instrument since different instruments probe different layers of
the atmosphere (see below). 
Therefore, one has difficulty to compare values of $H \,C_{\rm obs} $ coming from different instruments.


\subsection{Solar seismic constraints}

\citet{Baudin05} have inferred the solar $p$-mode excitation rates
from different instruments, namely GOLF on-board SOHO, the BiSON and GONG networks. 
As pointed out by \citet{Baudin05}, the layer
($r_h$) where the mode mass is evaluated must be properly estimated
to derive  correct values of the excitation rates from \eq{pow_obs}.  Indeed solar
seismic observations in Doppler velocity are usually measured 
from a given spectral line. The layer where the oscillations are measured then
depends on the height ($r_h$) in the atmosphere where the line is formed. Different
instruments use different solar lines and then probe  different
regions of the atmosphere.  For instance, the BiSON instruments use the KI line whose
height of formation is estimated at the optical depth  $\tau \approx
0.013$. The optical depth associated with the different spectral lines used in
helioseismology are given in \citet{Houdek06} with associated references.
 
\begin{figure}
\begin{center}
 \includegraphics[width=10cm]{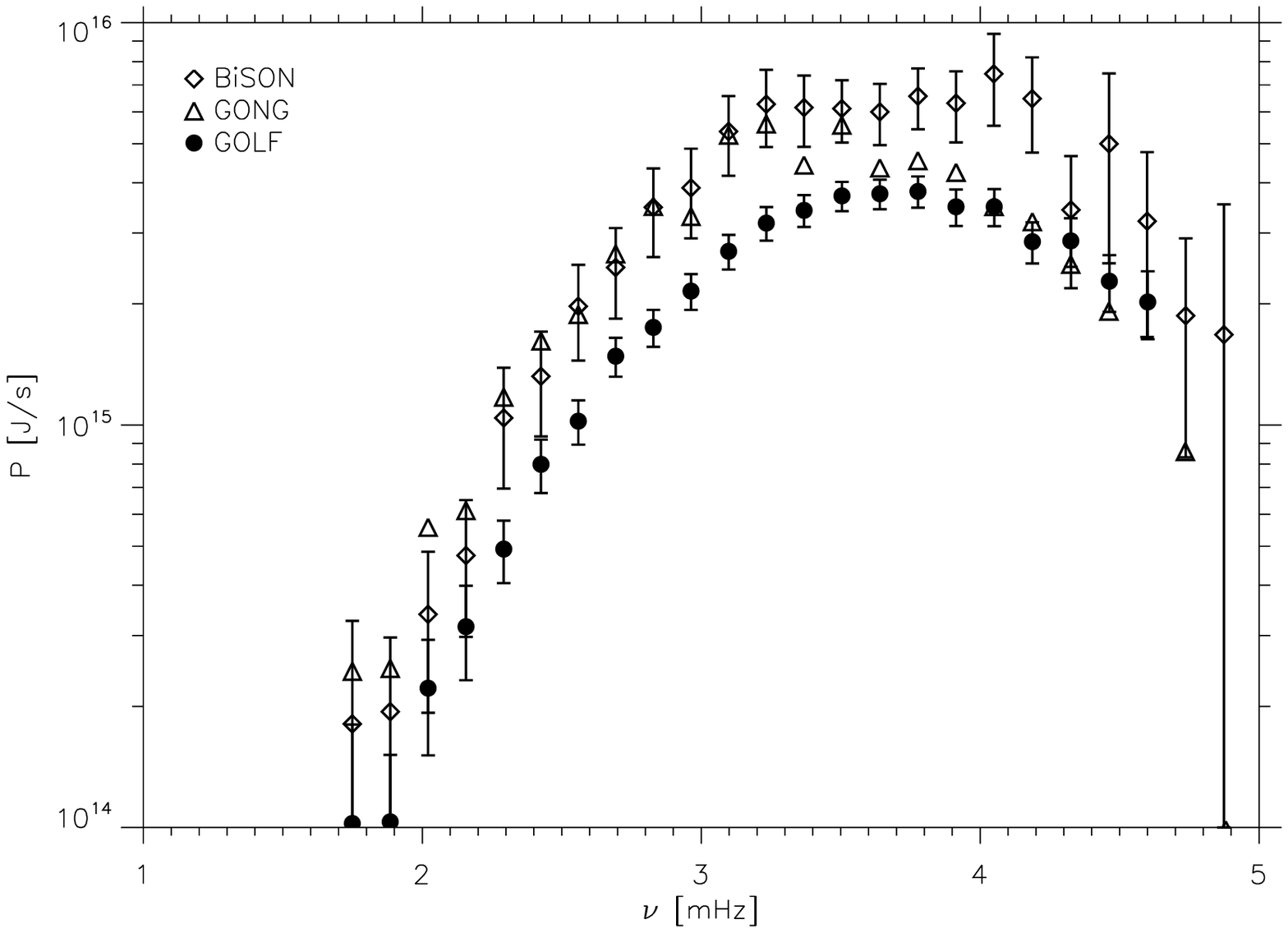} 
 \includegraphics[width=10cm]{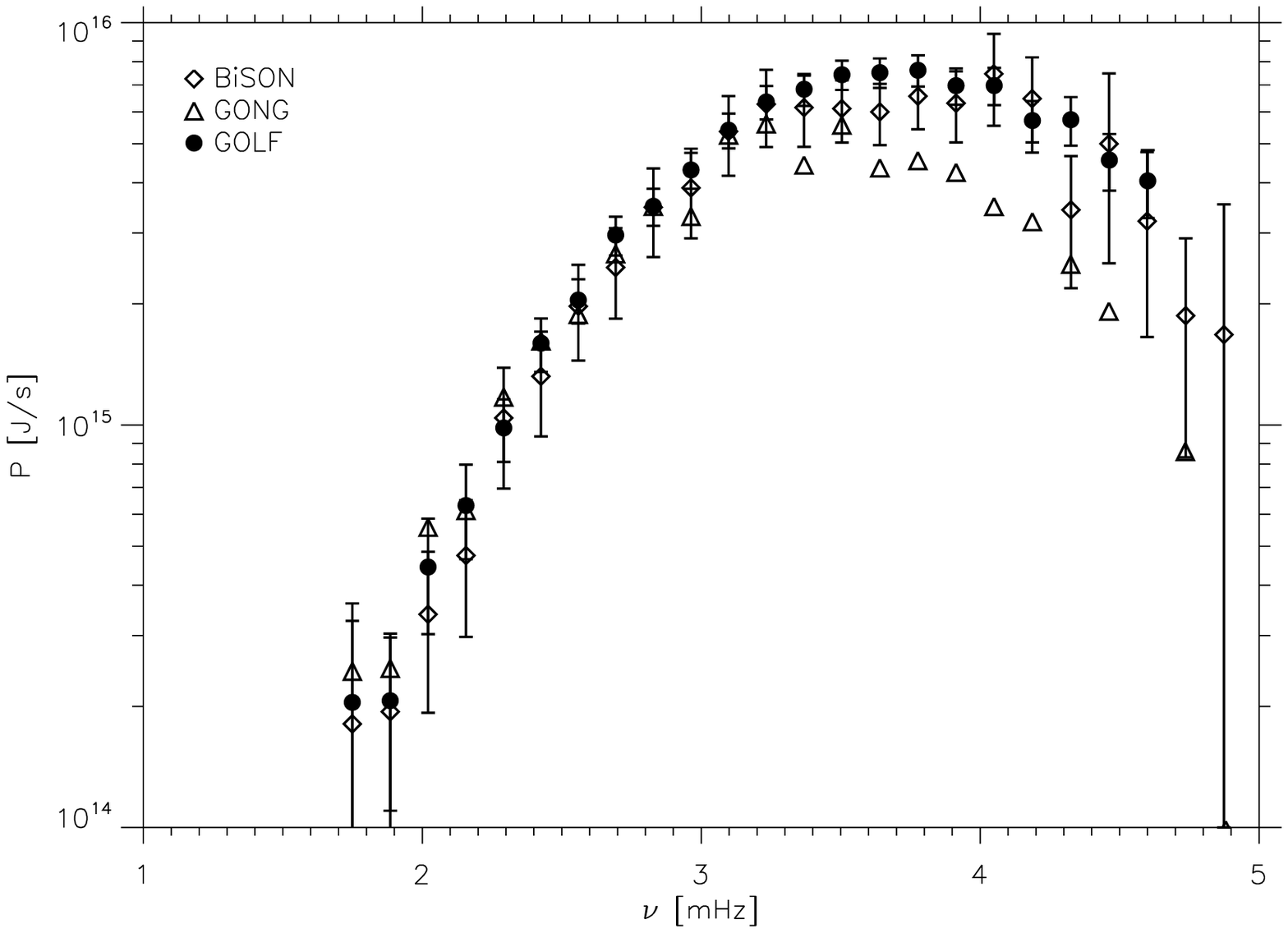} 
\end{center}
 \caption{{\bf Left:} Solar $p$-mode excitation rates, ${\cal P}$, as a function
   of frequency and derived from different instruments.  The filled
   circles correspond to seismic data from
   SOHO/GOLF, the diamonds to seismic data from the BiSon network, and the
   triangles to seismic data from the GONG network.
{\bf Right:} Same as left panel.  ${\cal P}$ derived from GOLF data 
multiplied by a factor in order that they match at low frequency
the ${\cal P}$ derived from GONG or BiSON.
}
\label{pow_sun}
\end{figure}

Solar $p$-mode excitation rates, ${\cal P}$, derived by \citet{Baudin05} are  shown
in Fig.~\ref{pow_sun} (left panel).  For $\nu \lesssim $~3.2~mHz, ${\cal P}^{\rm GONG}$ and ${\cal P}^{\rm
  BiSON}$ are consistent with each other, whereas ${\cal P}^{\rm
  GOLF}$  is systematically smaller than  ${\cal P}^{\rm GONG}$ and ${\cal P}^{\rm
  BiSON}$, although the discrepancy remains within 1-$\sigma$. 
At high frequency, differences between the different data sets are
more important. This can be partially attributed to the choice of the layers $r_h$ where
${\cal M}$ are evaluated. 
Indeed, the sensitivity of ${\cal M}$  to  $r_h$ is the larger at high
frequency.  
On the other hand, low-frequency mode masses
are much less sensitive to the choice of  $r_h$. Accordingly, the discrepancy
seen at low frequency between GOLF and the other data sets suggests
that the absolute calibration of the GOLF data may  not be correct
\citep[see][]{Baudin05}.  
In Fig.~\ref{pow_sun}, we then present  ${\cal P}$ derived from GOLF data
after multiplying them by a factor in order that they match at low
frequency  ${\cal P}^{\rm GONG}$ and ${\cal P}^{\rm  BiSON}$.
We find a rather good agreement between ${\cal P}^{\rm GOLF}$ and
${\cal P}^{\rm  BiSON}$ whereas, at high frequency, ${\cal P}^{\rm GONG}$ are
systematically lower than  ${\cal P}^{\rm GOLF}$ ore
${\cal P}^{\rm  BiSON}$. 
The residual high-frequency discrepancy is likely due to an incorrect
determination of the layer $r_h$ where the different seismic
measurements originate \citep[see a detailled discussion in][]{Baudin05}. 

\subsection{Stellar seismic constraints}
\label{stellar constraints}

\begin{figure}
\begin{center}
 \includegraphics[width=10cm]{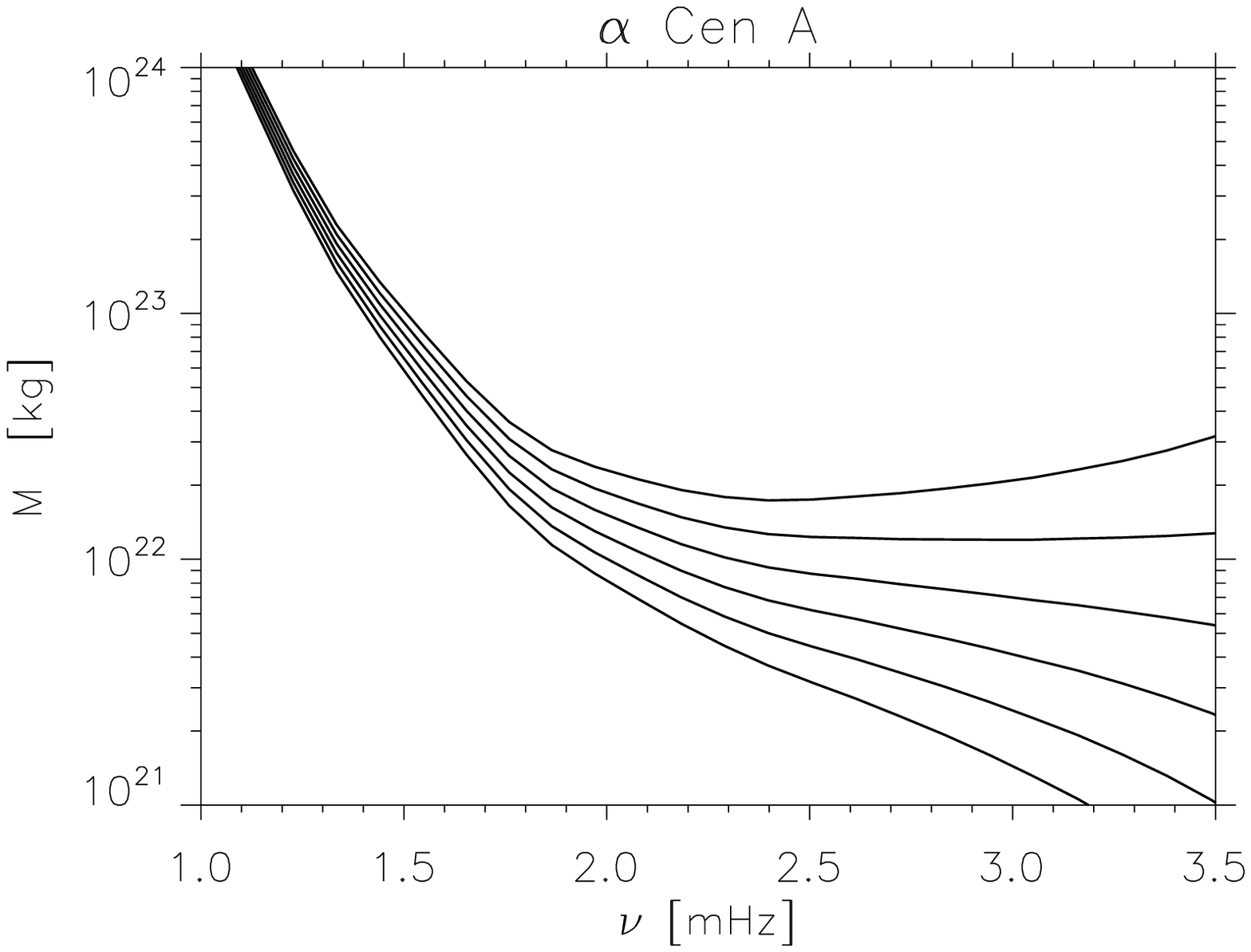} 
\end{center}
\caption{ Mode mass evaluated for the case of {\acenA} at  different heights $h$ above
the photosphere. The upper curve corresponds to the photosphere
($h=0$) and the lower curve to the top of the atmosphere ($h=$~1000~km). The step
in $h$ is 200~km. 
}
\label{MM_vs_h}
\end{figure}

Seismic  observations in Doppler velocity of solar-like pulsators are performed using 
spectrographs dedicated to stellar seismic measurements (e.g.  UCLES, UVES, HARPS). Such spectrographs use  a large number of spectral lines in order to reach a high 
enough signal-to-noise ratio. 
In the case of stellar seismic measurements, it is then  more difficult than
for helioseismic observations to estimate the effective height $r_h$.
As discussed in detail in \citet{Samadi08}, the computed mode surface
velocities, $v_s$,  depend significantly on the choice of the height, $h$,
in the atmosphere where the mode masses are evaluated.   
This is illustrated in Fig.~\ref{MM_vs_h} for the case of the star $\alpha$~Cen~A.

A recent work by \citet{Kjeldsen08} allows us to estimate the value for an effective $r_h$. 
Indeed, the authors have found that solar modes measured with the UCLES
spectrograph have amplitudes slightly smaller than those measured by
the BiSON network.  
The instruments of the BiSON network use the potassium (K)
resonance line, which is formed at an optical depth $\tau_{\rm 500~nm}
\simeq 0.013$.
\citet{Kjeldsen08}'s results then suggest that acoustic modes measured
by UCLES or an equivalent spectrograph (e.g. HARPS) are 
measured at an effective height ($r_h$) slightly below the formation depth of the K line,  i.e.
at optical depth slightly above  $\tau_{\rm 500~nm} \simeq 0.013$. 
Accordingly, in the case of stellar seismic observations we will
evaluated  the mode masses at that optical depth.  
A more rigorous approach would be  to compute an effective mode mass by
  weighting appropriately the different mode masses associated with the different
  spectral lines that contribute to the seismic measure.
In order to infer accurate mode excitation rates from
the stellar seismic data, the mode masses
representative of the observation technique  \emph{and} the 
  spectral lines of the observed star must be derived. However, this
  calls for further studies.

\section{Theoretical models}
\label{Theoretical models}

\subsection{The inhomogeneous wave equation}

Most of the theoretical models of stochastic excitation adopt
GK's approach. 
It consists to  solve first, with appropriate boundary
conditions, the equation that governs the \emph{adiabatic} wave
propagation (also called the homogeneous wave equation). This provides the
well-known adiabatic displacement eigenvectors (${\vec \xi}(\vec
r,t$)). Then, we include in the wave equation of propagation turbulent
sources of driving as well as a linear damping. 
The complete equation (so-called inhomogeneous wave
equation) is then  solved  and the solution corresponds to the forced mode
displacement, $\delta \vec r_{\rm osc} (\vec r, t)$ (or equivalently the
oscillation mode velocity $\vec
{v}_{\rm osc}   = d\delta r_{\rm osc}/dt$).  

A detailed derivation of the solution can be found in 
   \citet[][SG herefater]{Samadi00I} or in \citet{Chaplin05}. 
We recall below the main steps.

Equilibrium quantities are represented with a subscript 0.
 Each variable $f$, except for the velocity $\vec v$,  is written   as 
the sum of the equilibrium quantity , $f_0$ and an Eulerian  fluctuation, $f_1$,
$f = f_0 + f_1$ and 
we retain terms which are linear and quadratic in the variables 
$P_1$ and $\rho_1$ and  neglect, $g_1$, the gravitational perturbation
\footnote{Neglecting the perturbation of the gravity corresponds to
  \citet{Cowling41}'s approximation. This approximation remains valid
  for modes with a high $n$ radial order.}. Accordingly, one  obtains
for the perturbed momentum and continuity equations:
\eqna{
\derivp{\rho \vec v  } {t}  + \vec \nabla : ( \rho \vec v \vec v ) + 
\vec \nabla P_1 - \rho_ 1 \vec g_0   = 0
\label{perturbed_momentum_eqn}
\\
\derivp{ \rho_1} {t} + \vec \nabla . (  \rho \vec v) = 0 \; .
\label{perturbed_continuity_eqn}
}
where  $\omega_{\rm osc}$ is the mode frequency,  $P$, $\rho$,
 $\vec v$ and $\vec g$ denote respectively the gas pressure, 
density, velocity and gravity.

The perturbed  equation of state to second order in a Eulerian description is given by:
\eqna{
P_1 = c_s^2 \rho_1 +    \alpha_s  s_1 + \alpha_{\rho \rho} \rho_1 ^2
 + \alpha_{ss}  s_1^2 + \alpha_{\rho s} \rho_1 s_1
\label{perturbed_state_eqn}
}
where  $s$ is the entropy, $\displaystyle{\alpha_s =\left ( \partial P
  /\partial s  \right )_\rho}$,  $\ds{c_s = \Gamma_1 \,  P_0/\rho_0}$
denotes the  average sound speed, $\displaystyle{\Gamma_1 = \left(
  \partial \ln P / \partial \ln \rho \right )_s }$ is the  adiabatic
exponent and $\alpha_{\rho \rho} $, $\alpha_{ss}$ and $\alpha_{\rho
  s}$ are the second partial derivatives of $P$ versus $s$ and
$\rho$. Note that \eq{perturbed_state_eqn} assumes a constant chemical
composition (this is indeed the case in the outer convective layers)  but also 
constant ionisation rates.

The velocity field $\vec{v} $ is split into a component due to the  
pulsational displacement $\delta \vec r_{\rm osc} $  and a 
 turbulent  component $\vec{u}$  as
\eqn{
\vec{v} = \vec v_{\rm osc}  +  \vec{u}
\label{vosc+vturb}
}

Linearisation of Eq.(\ref{perturbed_momentum_eqn}-\ref{perturbed_state_eqn}) yields for the velocity field, 
in the absence of turbulence ($\vec u = 0$), the homogeneous wave equation
\eqna{
\left ( \derivp  { ^2 } {t^2}  -  \vec L  \right )  
\vec {v}_{\rm osc}   = 0
\label{homogeneous_wave}
}
where $L$ is the linear wave  operator (see its expression in SG).
With  appropriate boundary conditions \citep{Unno89}
one recovers 
 the usual eigenvalue problem~:
\eqn{
\vec L(\vec \xi(\vec r, t))  =  - ~ \omega_{\rm osc}^2 \; \vec \xi(\vec r, t)
}
where  $\omega_{\rm osc}$ is the mode eigenfrequency and $\vec \xi(\vec r,t)
\equiv e^{-i \omega_{\rm osc} t} \, \vec{\xi} (\vec {r})$  is the
adiabatic displacement  eigenvector.

In the presence of turbulence, the pulsational displacement ($\delta
\vec r_{\rm osc}$)  is written  in terms of the above adiabatic 
solution $\vec {\xi} (\vec{r},t)$  and an instantaneous  
amplitude $A(t)$ according to \eq{delta_osc_2}.  
Under the assumption of a slowly varying intrinsic amplitude $A(t)$,
the velocity ($\vec v_{\rm osc}$) is related to  $A(t)$ and $\delta
\vec r_{\rm osc}$ according to \eq{At}.

Differentiating Eq. (\ref{perturbed_momentum_eqn}) 
with respect to $t$, subtracting the time averaged equation of
motion, neglecting  non-linear terms in $\vec v_{\rm osc}$,
 assuming an incompressible turbulence ($\vec \nabla . \vec u =0$)   
and using
Eqs. (\ref{perturbed_continuity_eqn}) and (\ref{perturbed_state_eqn})
 yields  the inhomogeneous wave equation 
\eqna{
\rho_0 \left ( \derivp  { ^2 } {t^2}  -  \vec L  \right )
\left [ \vec v_{\rm osc}  \right ] +
\vec {\cal D} \left [ \vec v_{\rm osc}\right ]  & = &  \derivp{}{t} \vec {\cal S} - \vec {\cal C}
\label{inhomogeneous_wave}
}
with 
\eqna{
  \vec {\cal S} & \equiv & \vec {\cal S}_R + \vec {\cal S}_S \\
\vec {\cal S}_{R} & = & \vec \nabla : \left (\rho_0 \, \vec u \, \vec u
\right ) -  \vec \nabla : \left ( \left < \rho_0 \, \vec u \, \vec u \right >
\right )
\label{S_R}
\\
\vec {\cal S}_{S} & = & - \vec \nabla \, \left ( \bar{\alpha}_s \, s_t \right )
\label{S_S}
}
where $s_t$ is the \emph{Eulerian} turbulent entropy fluctuations and
$\overline \alpha_s = \overline{ (\partial P /  \partial \rho)_s }$.
The terms ${\cal S}_R$ (\eq{S_R}) and ${\cal S}_S$ (\eq{S_S}) are two driving
  sources, namely the Reynolds stress tensor and a term that involves
  the Eulerian entropy fluctuations. 
The last term ${\cal C}$ in the RHS of \eq{perturbed_continuity_eqn}
gathers terms that involve $\rho_1$ as well as the second order
terms of \eq{perturbed_state_eqn}. ${\cal C}$ can in principle contribute to the
driving.  However, one can show that its 
contribution is negligible compared to ${\cal S}_R$  and ${\cal S}_S$ (see SG,
GK).

The operator $\vec {\cal D}$ in the LHS of
\eq{perturbed_continuity_eqn} involves both the turbulent velocity field
($\vec u$)  and the pulsational velocity.  This term contributes to
the  dynamical linear damping.

As we will see later, it is more convenient to decompose the Eulerian
entropy fluctuations in terms of the Lagrangian ones, that is as:
\eqna{
{ {\partial s_t} \over {\partial t}}   & = &  { { d \delta s_t } \over {dt} }  - \vec u
\, . \nabla (s_0 + s_t )
}
where $s_0$ is the mean entropy. 
Accordingly, $S_S$ is such that:
\eqna{
{ {\partial \vec {\cal S}_S } \over {\partial t}}  & = & - \vec \nabla \, \left ( {d \over
  {dt} } \left ( \bar{\alpha}_s \, \delta s_t  \right ) - \bar{\alpha}_s \, \vec u
\, . \vec \nabla \,  s_t  \right ) \,
\label{S_S_2}
}
where we have dropped the term $\vec u \, . \vec \nabla s_0$ since it
does not contribute to the driving (GK, see also SG).
Integration of Eq.~(\ref{S_S_2}) with respect to time  then gives
${\cal S}_S$.

\subsection{General solution}

Substituting Eq. (\ref{At}) into
 Eq. (\ref{inhomogeneous_wave}), yields, with the 
help of Eq. (\ref{homogeneous_wave}),  a differential equation for $A(t)$. This latter equation is straightforwardly solved  and one obtains the solution for $A$:
\eqn{
A(t)  =  \frac {i e^{-\eta t} } {2 \omega_{\rm osc} I}  \int_{-\infty}^{t} dt^\prime   
\int_{\cal V} d^3 x \,  e^{(\eta+i\omega_{\rm osc}) t^\prime} \,  
 \vec{\xi}^* (\vec x).  \vec {\cal S }  (\vec x, t^\prime) 
\label{A_t} 
}
where $I$ is the mode inertia (which expression is given in
\eq{inertia}) and the spatial integration is performed over the
stellar volume, $\cal V$.
As the sources are random, $A$ can only be calculated  in square
average, $\langle |A|^2 \rangle$. This statistical average is performed over a large set of
realizations. 
From Eq.\ (\ref{A_t}) and with the help of some  simplifications as detailed in SG, one finds:
\eqna{ 
 \left < \left | A
   \right | ^2 \right >  & =  &    \frac{C^2  }{8 \, \eta \,( \omega_{\rm osc}\,   I)^2} \,  \label{A2} \; ,
}
with
\eqna{ 
C^2 & \equiv &  \int_{{\cal V}}  d^3x_0 \int_{-\infty}^{+\infty} d^3r \, d\tau \, 
e^{-i\omega_{\rm osc} \tau}  \left <     \vec \xi^* \,  .\, \vec {\cal{S}} _1  \,  
 \vec \xi \,   .  \,   \vec {\cal S}_2  \right > \label{C2} 
\label{S}
}
where  $\eta$ is the mode damping rate (which can be derived from
seismic data), $I$  the mode inertia (\eq{inertia}), $\vec x_0$
 the position in the star where the stochastic excitation is
 integrated, ${\cal V}$ is the volume of the convective region, ${\cal S}$ represents the different driving terms,
$\vec r$,
 and $\tau$ are the spatial correlation and temporal correlation lengths
 associated with the local turbulence, subscripts 1 and 2
 refer to quantities that are  evaluated at the spatial and
 temporal positions $[ \vec x_0-\frac{\vec r}{2}, - \frac{\tau} {2}]$  
and $ [\vec x_0+\frac{\vec r}{2}, \frac{\tau} {2}]$
respectively, and finally $\langle . \rangle$ refers to a statistical average.  

According to \eqs{balance_3}, (\ref{E_osc_2}) and (\ref{A2}), the theoretical mode excitation rate, ${\cal P}$,
is then given by the  expression:
\eqna{
{\cal P} & = & \frac{C^2 } {8 \, I }
\label{pow_1}
}

\subsection{Driving sources}
 \label{Driving sources}

The Reynolds stress tensor (Eq.~(\ref{S_R})) was identified early on by \citet{Lighthill52}
as a source of acoustic noise and then as a source of \emph{mode} excitation
(GK). 
This term represents a mechanical source of driving and is considered by most of the theoretical
formulations as the dominant contribution to the mode excitation
\citep{GK77,Dolginov84,Balmforth92c,Stein01II,Samadi02II,Chaplin05}.
However, as pointed-out by \citet{Osaki90}, the first calculations by GK's 
significantly under-estimate the power going to the solar modes compared
to the observations. 

In order to explain the mode excitation rates derived from the
observations, \citet[][GMK hereafter]{GMK94} identified the
\emph{Lagrangian} entropy fluctuations, i.e. the term $\delta s_t$ in
Eq.~(\ref{S_S_2}), as an additional driving source.
These authors claimed that this term is the dominant source of driving.
However, GMK assumed that entropy fluctuations ($s_t$) behave as a
passive scalar. A passive scalar $f$ is a quantity that
obeys an equation of diffusion \citep[see e.g.][]{Lesieur97}:
\eqna{
{ {d f } \over {dt} }   & = & { { \partial f} \over {\partial t}}  + \vec u
\, . \vec \nabla \, f  = \chi \, \nabla^2 \,  f   \; ,
\label{scalaire_passif}
}
where $\chi$ is a diffusion coefficient.  As shown by SG, 
assuming as GMK that  $\delta s_t$  is a passive scalar leads to a vanishing
contribution. On the other hand, SG have shown that the term
$\bar{\alpha}_s \, \vec u \, . \vec
\nabla \, s_t$  in the RHS of Eq.~(\ref{S_S_2}) contributes
effectively  to the mode driving. 
In SG formulation, the so-called entropy source
term is then~:
\eqna{
{\partial \over {\partial t}} {\cal S}_S & = &  \vec \nabla \, \left (
\bar{\alpha}_s \, \vec u
\, . \vec \nabla \,   s_t \right )  \; .
\label{S_S_3}
}
The term $\vec u\, . \vec \nabla \,   s_t$ in the RHS of \eq{S_S_3}
is an advective term. 
Since it involves the entropy fluctuations it
can be considered as a thermal source of driving.  
The source term of \eq{S_S_3} was also identified by GK, but was
considered as negligible. 
It must also be pointed out that the theoretical formalisms by \citet{Balmforth92c} and \citet{Chaplin05}
 did not consider this source term. 
According to \citet{Samadi02II}, this term is not negligible (about $\sim$ 15~\% of
the total power) but nevertheless small
compared to the Reynolds stress source term (${\cal S}_R$) in the case
of the Sun.

Finally, as seen in Eq.~(\ref{S}), ${\cal S_R}$ and ${\cal S}_S$
lead to cross terms. 
However, assuming as GMK that $s_t$ behaves as a passive
scalar and an \emph{incompressible} turbulence (i.e. $\vec
\nabla .  \, u=0$),   SG have shown that the crossing term between ${\cal S_R}$ and
${\cal S}_S$ vanishes. Hence, in the framework of those assumptions, 
there is \emph{no canceling} between the two contributions (but see Sect.~\ref{entropy}).

\subsection{Length scale separation}
\label{Length scale separation}

As seen in the RHS of \eq{S}, the eigen-displacement $\vec \xi(\vec
r)$ is  coupled spatially with the source function, ${\cal
  S}$.  In order to derive a theoretical formulation that can be
evaluated, it is necessary to spatially decouple   $\vec \xi(\vec
r)$ from  ${\cal S}$. This is the reason why all theoretical
formulations explicitly or implicitly assume that 
eddies that effectively contribute to the driving have a characteristic
length scale  smaller than the mode wavelength.
Indeed, provided this is the case,  $\vec \xi(\vec
r)$ can be  removed  from the integral over $r$ and
$\tau$ that appears in the RHS of Eq.~(\ref{A2}) (see SG). 
This  assumption is justified for low turbulent Mach numbers $M_t$ ($M_t \propto u
/ c_s$ where $c_s$ is the sound speed).  
However, at the top of the solar convective zone, that is in the
super-adiabatic region, $M_t$ is no longer small ($M_t \sim 0.3$).  
Furthermore, for G and F stars lying on the main sequence, $M_t$ is expected to increase
with the effective temperature  and to reach a maximum for $M \sim 1.6 \, M_\odot$
\citep[see][]{Houdek99}.  Hence, for F type stars, significantly hotter
than the Sun, the length scale separation becomes a more questionable
approximation (see the discussion in Sect.~\ref{discussion}).

\subsection{Closure models}
\label{Closure models}

The second integral in RHS  of Eq.~(\ref{A2}) involves the  term $\left < {\cal S}_1 \, {\cal S}_2 \right
>$, which is a two-point spatial \emph{and} temporal correlation products of the source
terms. Hence, the Reynolds stress source term (Eq.~(\ref{S_R})) leads to
the two-point  correlation product of the form $ \langle (\vec u \, \vec u )_1 \, (\vec u 
\, \vec u )_2  \rangle $. In the same  way, the entropy source term
(Eq.~(\ref{S_S_3}))  leads to the two-point  correlation product of the form $ \langle (\vec u \, s_t )_1 \, (\vec u 
\, s_t )_2 \rangle$. In both case, we deal with fourth-order two-point
correlation product  involving
turbulent quantities (that is $\vec u$ and $s_t$). Fourth-order
moments are solutions of equations involving fifth-order moments. In
turn, fifth-order moments are expressed in term of six-order moments
... and so on.  
This is the well known closure
problem.  A simple closure model is the quasi-normal approximation
(QNA hereafter) that permits one to express fourth order moments in term
of second order ones \citep[see details in e.g.][]{Lesieur97}, that is :
\eqna{
 \langle (u_i \, u_j )_1 \, (u_k  \, u_l  )_2 \rangle  (\vec r,\tau)  = & \langle (u_i \, u_j )_1 \rangle
 \, \langle  (u_k \, u_l )_2 \rangle  +   \, \langle ( u_i )_1 \, (u_l)_2\rangle \,  \langle ( u_j )_1 \,
 (u_k)_2\rangle      \nonumber  \\  &  + \, \langle ( u_i )_1 \, (u_k)_2\rangle \,  \langle ( u_j )_1 \,
 (u_l)_2\rangle  &  \; 
\label{qna}
}
The decomposition of \eq{qna} is strictly valid when the velocity is normally
distributed.
The first term in the RHS of Eq.~(\ref{qna}) cancels  the term $
\left < \vec u \, \vec u \right >$ in Eq.~(\ref{S_R}) \citep[see
 details in][]{Chaplin05}. An expression similar to
Eq.~(\ref{qna}) is derived for 
the correlation product $ \langle (\vec u \, s_t )_1 \, (\vec u  \, s_t )_2
\rangle $ (see SG).  

\subsection{Adopted model of turbulence}
\label{Adopted model of turbulence}

It is usually more convenient to express \eq{qna} in the frequency
($\omega$) and wavenumber ($k$) domains.
We then define $\phi_{i,j}$ as the temporal and spatial Fourier transform of
$ \langle ( u_i )_1 \, (u_j)_2 \rangle$. For an inhomogeneous, incompressible, isotropic and
stationary turbulence, there is a relation between $\phi_{i,j}$ and
the kinetic energy spectrum $E$, which is \citep{Batchelor70}:
\eqna{
\phi_{ij}( \vec k,\omega) &  =& \frac { E( k,\omega) } { 4 \pi k^2}  \left( \delta_{ij}- \frac {k_i k_j} {k^2}  \right)
\label{phi_ij}
} 
where $k$ and $\omega$ are the wavenumber and frequency respectively
associated with the turbulent elements, and  $\delta_{i,j}$ is the
Kronecker symbol.  Following \citet{Stein67}, it is possible to
split for each layers $E(k,\omega)$ as:
\eqna{
E(k,\omega) & = & E(k) \, \chi_k(\omega) 
\label{ek_chik}
}
where $E(k)$ is the time averaged kinetic energy spectrum and
$\chi_k(\omega)$ is the frequency component of $E(k,\omega)$. In other
words, $\chi_k(\omega) $  measures -~  in the frequency and $k$ 
wavenumber  domains ~-  the temporal correlation between eddies. 
As discussed in Sect.~\ref{Eddy time-correlation}, the way the eddy time-correlation is modeled has an important consequence on the efficiency of the mode driving. 
A decomposition similar to that of \eq{ek_chik} is performed for the
spectrum associated with the entropy fluctuations ($E_s(k,\omega)$). 

Note that $\chi_k(\omega)$ and $E(k)$ satisfy by definition the
following normalisation conditions: 
\eqna{
\int_{-\infty}^{+\infty} d\omega \,  \chi_k (\omega)  = 1 \; , \\
\label{eqn:chi_omega_norm}
\int_0^{\infty}{\rm d}k\,E(k)  =  \ds  {1 \over 2} \, {\langle \vec u^2 \rangle  } =
{ \Phi \over 2 }\, \langle u_z^2 \rangle \equiv   \ds {3 \over 2} \,  u_0^2 \; ,
\label{eqn:E:normalisation}
}
where $u_z$ is the vertical component of the velocity, $\Phi \equiv  \langle u^2 \rangle / \langle u_z^2 \rangle $ is the anisotropy factor introduced
by \citet{Gough77}, and $u_0$ is a characteristic velocity introduced
for convenience.
A normalisation condition similar to \eq{eqn:E:normalisation} is
introduced for  $E_s(k)$ (see details in SG).

\subsection{Complete formulation}

On the basis of the different assumptions mentioned above, SG then
derive for  \emph{radial} modes the following theoretical expression
for ${\cal P}$ :
\eqna{ {\cal P}  & = & \frac{1}{8 \,  I } \left ( C_R^2 +  C_S^2 \right )
\label{pow_2}
}
where $C_R^2$ and $C_S^2$ are the turbulent Reynolds stress and entropy contributions
respectively. There expressions are (see SG):
\eqna{ 
C_R^2 & = &  4 \, \pi^{3} \, {\cal G} \, \int_{0}^{M}
{\rm d} m \, \rho_0 \left |  \deriv { \xi_{\rm r}} {r} \right | ^2 \, S_R(m,\omega_{\rm osc})
\label{C2R} \\
C_S^2 & = &   \frac{ 4 \, \pi^3 \, {\cal H} } {\omega_{\rm osc}^2} \,
\int_{0}^{M} {\rm d} m  \,  { \bar{\alpha}_s^2 \over \rho_0 } \,
g_{\rm r} (\xi_{\rm r},m) \, S_S(m,\omega_{\rm osc})  
\label{C2S} 
}
with $S_R$ and $S_S$ are the source terms associated with the Reynolds
stress and entropy fluctuations respectively:
\eqna{
S_R &  = &\int_0^\infty dk \,
 \frac {E^2(k,m)} { k^2} \, \int_{-\infty}^{+\infty}  d\omega \, \chi_k (
\omega_{\rm osc} + \omega, m) \, \chi_k (\omega, m) \label{SR}
\\
S_S &  = & \int_{0}^\infty  dk \,
  \frac { E_s(k,m) E(k,m) } { k^2 } \, \int_{-\infty}^{+\infty} d\omega \,
 \chi_k(\omega_{\rm osc}+ \omega,m) \chi_k(\omega,m) \;
\label{SS}
}
In Eq.\ (\ref{C2R}) and (\ref{C2S}),  $\rho_0$ is the mean density,
${\cal G}$ and ${\cal H}$ are two anisotropic factors 
(see their expressions in SG), and finally $g_{\rm r} (\xi_{\rm r},m)$ is a function that involves the
first and the second derivatives of $\xi_{\rm r}$, its expression is:
\eqn{
 g_{\rm r}(\xi_{\rm r},m) = 
\left( {1 \over\alpha_s } \deriv{\alpha_s }{r} 
  \,  \deriv { \xi_r} {r}  - \deriv{^2 \xi_r } {r^2} \right )^2
}
It is in general more convenient to rewrite Eqs.~(\ref{C2R}) and
(\ref{C2S}) in the following forms:
\begin{eqnarray}
\label{C2R_1}
C_R^2 & =  & { 4 \pi^{3} \mathcal{G}} \int_0^M {\rm d} m  \, {   \rho_0 \,  u_0^4  \over
  {k_0^3  \, \omega_0} }   \,  \left |  \deriv { \xi_{\rm r}} {r}
\right | ^2 \, \tilde{S}_R(m,\omega_{\rm osc}) \; ,\\
\label{C2S_1}
C_S^2 &=& \frac{4 \pi^3 \mathcal{H} }{\omega_{\rm osc}^2}  \int_0^M
{\rm d} m  \, {
    { (\bar{\alpha}_s \, \tilde{s} \,  u_0)^2 }\over { \rho_0 \, k_0^3 \,
      \omega_0 }  } \,  g_{\rm r } (\xi_{\rm r},m) \, \, \tilde{S}_s(m,\omega_{\rm osc})
\end{eqnarray}
where we have defined the dimensionless source functions $\tilde{S}_R
\equiv \left ( k_0^3 \, \omega_{0} \, / \, u_0^4 \right ) \, S_R $ and
$\tilde{S}_s \equiv  \left ( k_0^3 \, \omega_{0} \, / \, ( u_0^2 \,  \tilde{s}
² )  \right ) \, S_R $, $\tilde{s}$ and where $\tilde s$ is the rms of
the entropy fluctuations.
We have introduced for convenience  the
characteristic frequency $\omega_0$  and  the characteristic wavenumber
$k_0$ ; they are defined as:
\eqna{
 \omega_0 & \equiv & k_0 \,  u_0  \label{omega_0}\\
 k_0 &\equiv &{ {2 \pi} \over \Lambda} \label{k_0}   \label{k0}
}
where  $\Lambda$ is a characteristic size derived from $E(k)$ and
$u_0$ is the characteristic velocity given by
\eq{eqn:E:normalisation}. 
For future use, it is also convenient to define a characteristic time $\tau_0$ as:
\eqn{
\tau_0 = { { 2 \pi } \over {k_0 \, u_0 } }  = { \Lambda \over u_0}
\label{tau_0}
}


From \eq{C2R_1} we can show that the driving by the Reynolds stress
is  locally proportional to the kinetic energy flux. Indeed, the flux
of kinetic energy  in the vertical direction is by definition:
\eqn{
F_{\rm kin} \equiv w \, E_{\rm kin} = w \, \left ( {1 \over 2} \, \rho_0 \, \vec{u}^2 \right )
= {3 \over 2} \, \sqrt{ 3 \over \Phi }  \, \rho_0 \, u_0^3 \; ,
\label{F_kin_z}
}
where $E_{\rm kin} \equiv (1/2) \, \rho_0 \, \vec{u}^2 $ is the kinetic
energy per unit volume.
Substituting \eq{F_kin_z} into \eq{C2R_1} yields the relation:
\eqn{ 
C_R^2  \propto  \int_0^M {\rm d} m  \, F_{\rm kin} \, \Lambda^4 \,  \left |  \deriv { \xi_{\rm r}} {r}
\right | ^2 \, \tilde{S}_R(m,\omega_{\rm osc}) \; .
\label{C2R_propto}
}
Concerning the  driving by the entropy
fluctuations, we can show that locally this driving does not only depend
on  $F_{\rm kin}$ but also on the convective flux ($F_c$).
Indeed, lets define  as GMK the quantity:
\eqn{
{\cal R} \equiv \frac{ \alpha_s   \tilde s  }  {\rho_0 u_0^2 } \;.
\label{R_GM}
}
Substituting \eq{R_GM} into \eq{C2S_1} yields the relation:
\eqn{ 
C_S^2  \propto  \int_0^M {\rm d} m  \, F_{\rm kin} \, \Lambda^4 \,
{\cal R}^2 \, {\cal F} ^2 \, \left ( { \omega_0 \over \omega_{\rm osc}
}  \right )^2\,\tilde{S}_S(m,\omega_{\rm osc}) \; ,
\label{C2S_propto}
}
where we have defined as in SG the quantity ${\cal F}^2 \equiv \Lambda^2 \, g_{\rm r}
$.  Finally, since  ${\cal R} \propto F_c / F_{\rm kin}$
\citep[see][]{Samadi05b}, we can conclude that locally the driving by
entropy source term is proportional to $F_{\rm kin}$ and to  the square of the ratio
${\cal R} \propto F_c/ F_{\rm kin}$.


\section{Turbulent spectrum}
\label{turbulent_spectrum}

As seen in Sect.~\ref{Adopted model of turbulence}, the model of
stochastic excitation developed by SG involves $E(k,\omega)$, the turbulent kinetic spectrum
 as well as $E_s(k,\omega)$, the spectrum associated with
the turbulent entropy fluctuations.
Both spectra are split in terms of a time averaged spectrum ($E(k)$ for
the velocity and $E_s(k)$ for the entropy fluctuations), and a
frequency component $\chi_k(\omega)$ (see Sect.~\ref{Adopted model of turbulence}).
Different prescriptions  were investigated for both components.
The results of these investigations are summarized in Sect.~\ref{Time
  averaged spectrum} for $E(k)$ and in Sect.~\ref{Eddy
  time-correlation} for  $\chi_k(\omega)$.

\subsection{Time averaged spectrum, $E(k)$}
\label{Time averaged spectrum}

Two approaches are commonly adopted for prescribing $E(k)$.
The classic one is to assume an analytical function derived
either from theoretical considerations or empirical ones. 
The more commonly used analytical  spectrum is the so-called
Kolmogorov spectrum \citep{Kolmogorov41}, which derives originally from
\citet{Oboukhov41}'s postulate that energy is transferred from the
large scales to the small scales at a constant rate. Other
theoretical spectra, such as the so-called Spiegel's spectrum
\citep{Spiegel62}, or purely empirical spectra, such as those proposed
by \citet{Musielak94}, were also considered.
All of these analytical functions differ from each other by the way
$E(k)$ varies with $k$. But for all of them, it is required to set a
priori the characteristic wavenumber, $k_0$, at which energy is injected
into the turbulent cascade.
The second approach consists to obtain $E(k)$ directly from hydrodynamical 3D
simulations. This method has two advantages: it provides both the  $k$
dependence of $E(k)$ and the characteristic wavenumber $k_0$.
On the other hand, the inconvenient is that such method depends on the quality of the 3D hydrodynamical simulation.

These two approaches have been compared in \citet{Samadi02I}. Among
the different analytical functions tested, the best 
agreement with a solar 3D simulation  was found with the so-called
``Extended Kolmogorov Spectrum'' defined by \citet{Musielak94}.  
This spectrum increases at low scales as $k^{+1}$ and decreases at low
scales according to the Kolmogorov spectum, i.e. as $k^{-5/3}$. 
However, due to the limited spatial resolution of the solar
simulation used,  the Kolmogorov scaling is validated over a
limited range only.  
Nevertheless, the major part of the excitation arises from the most-energetic
eddies, also refered to the energy bearing eddies. Accordingly, the
contribution of the small scales, that are not resolved by the present
3D simulations, are expected to be relatively small. However, to
confirm this, a quantitative estimate must be undertaken.

More important is the choice for the  characteristic wavenumber
$k_0$. Indeed, the integrands of \eq{C2R_1} and 
\eq{C2S_1} are both proportional to $k_0^{-4}$. Accordingly, the computed
${\cal P}$ are very sensitive to the choice for $k_0$. 
This characteristic wavenumber can be obtained from 3D simulations.
However, by default, one usually relates $k_0$ to the mixing-length
$\Lambda_{\rm MLT}$ according to:
\eqn{
k_0 = k_0^{\rm MLT} \equiv { { 2 \pi} \over {\beta \Lambda_{\rm MLT}} }
}
where $\Lambda_{\rm MLT} = \alpha \, H_p$ is the mixing-length, $\alpha$ the
mixing-length parameter, $H_p$ the pressure scale height, and 
$\beta$ a free parameter, which is usually set to a value of the order of
one. The solar 3D simulation used by \citet{Samadi02I} indicates that
in  the Sun $k_0 \simeq 3.6$~Mm$^{-1}$ at the top of the excitation
region. This characteristic wavenumber corresponds to an horizontal size of the granules of
$\Lambda_g = 2\pi/k_0 \sim~$2~Mm. This horizontal size is reached at
the top of the excitation region with a value of $\beta$ that depends
on the adopted value for $\alpha$ and the solar 1D model used. 
For other stars, 3D simulations are  rarely available. In that case,
one  usually assumes for $\beta$ the same value that the one adopted for the Sun.
Hence, an open and important question is whether or not the parameter $\beta$ can be
kept the same for other stars as for the Sun.
%

\subsection{Eddy time-correlation, $\chi_k(\omega)$}
\label{Eddy time-correlation}

\begin{figure}
\begin{center}
 \includegraphics[width=10cm]{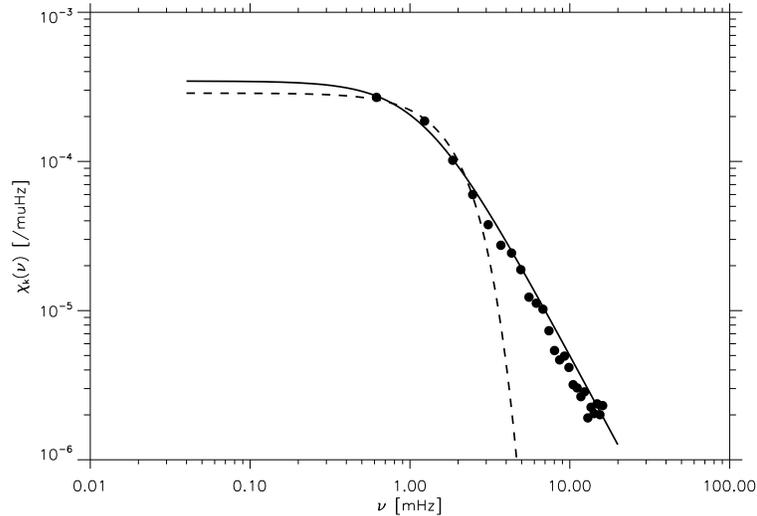} 
\end{center}
 \caption{ Eddy time-correlation function, $\chi_k$, as a function of
   frequency $\nu$ for the layer where the radial component of the velocity is
   maximum.  The  filled dots represent 
 $\chi_k$ obtained from a solar 3D simulation with an horizontal
   resolution of $\simeq$ 25~km \citep{Samadi02II}. $\chi_k$ is shown here
   for the wavenumber $k$ at which $E(k)$ peaks. 
The solid line represents a Lorentzian  function and the dashed line a
Gaussian function.
}
\label{kwpower}
\end{figure}

Most of the theoretical formulations explicitly or implicitly assume
a Gaussian function for $\chi_k(\omega)$
\citep[][]{GK77,Dolginov84,GMK94,Balmforth92c,Samadi00II,Chaplin05}. 
However, 3D hydrodynamical simulations of the outer layers
of the Sun show that, at the length associated with the energy
bearing eddies, $\chi_k$ is rather Lorentzian \citep{Samadi02II}. 
This is well illustrated in Fig.~\ref{kwpower}. 
As pointed-out by \citet{Chaplin05}, a Lorentzian  $\chi_k$  is also a
result predicted for the largest, most-energetic eddies by the
time-dependent mixing-length formulation derived by \citet{Gough77}. 
Therefore, there is some numerical and theoretical evidences that   $\chi_k$
is rather Lorentzian at the length scale  of the energy bearing
eddies.

\begin{figure}
\begin{center}
 \includegraphics[width=10cm]{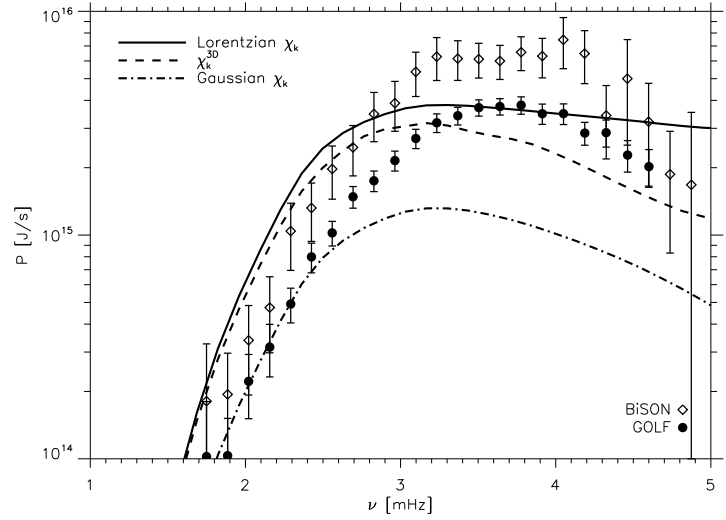} 
\end{center}
 \caption{ Solar $p$-mode excitation rates as a function of $\nu$. 
Filled circles and diamonds correspond as in Fig.~\ref{pow_sun} to seismic data 
from SOHO/GOLF and BiSON network respectively. The  lines correspond to
 semi-theoretical  calculations based on different choices for
 $\chi_k$: Lorentzian $\chi_k$ (solid line),  $\chi_k{\rm 3D}$ 
   i.e. $\chi_k$ derived directly from the
 solar 3D simulation (dashed line), and a Gaussian $\chi_k$ (dot-dashed line).
}
\label{pow_sun_2}
\end{figure}
\begin{figure}
\begin{center}
 \includegraphics[width=10cm]{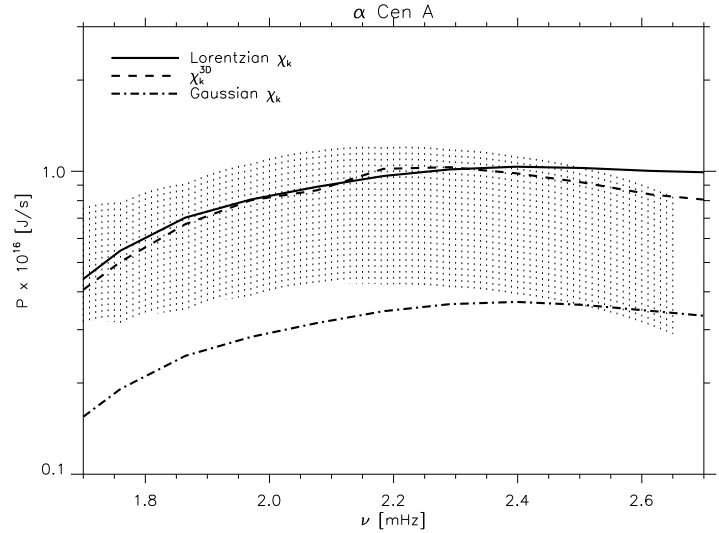} 
\end{center}
 \caption{Same as in Fig.~\ref{pow_sun_2} for the case of $\alpha$~Cen~A.
}
\label{pow_acena}
\end{figure}

As shown by \citet{Samadi02II}, calculation of the mode excitation
rates based on a Gaussian $\chi_k$ results for the Sun in a significant
under-estimation of the maximum of ${\cal P}$ whereas  a
better agreement  with the observations is found when a Lorentzian $\chi_k$ is
used. A similar conclusion is reached by \citet{Samadi08} in the case
of the star $\alpha$~Cen~A. These results are illustrated in Fig.~\ref{pow_sun_2}
in the case of the Sun and in Fig.~\ref{pow_acena} in the case of
$\alpha$~Cen~A. 

\begin{figure}
\begin{center}
 \includegraphics[width=10cm]{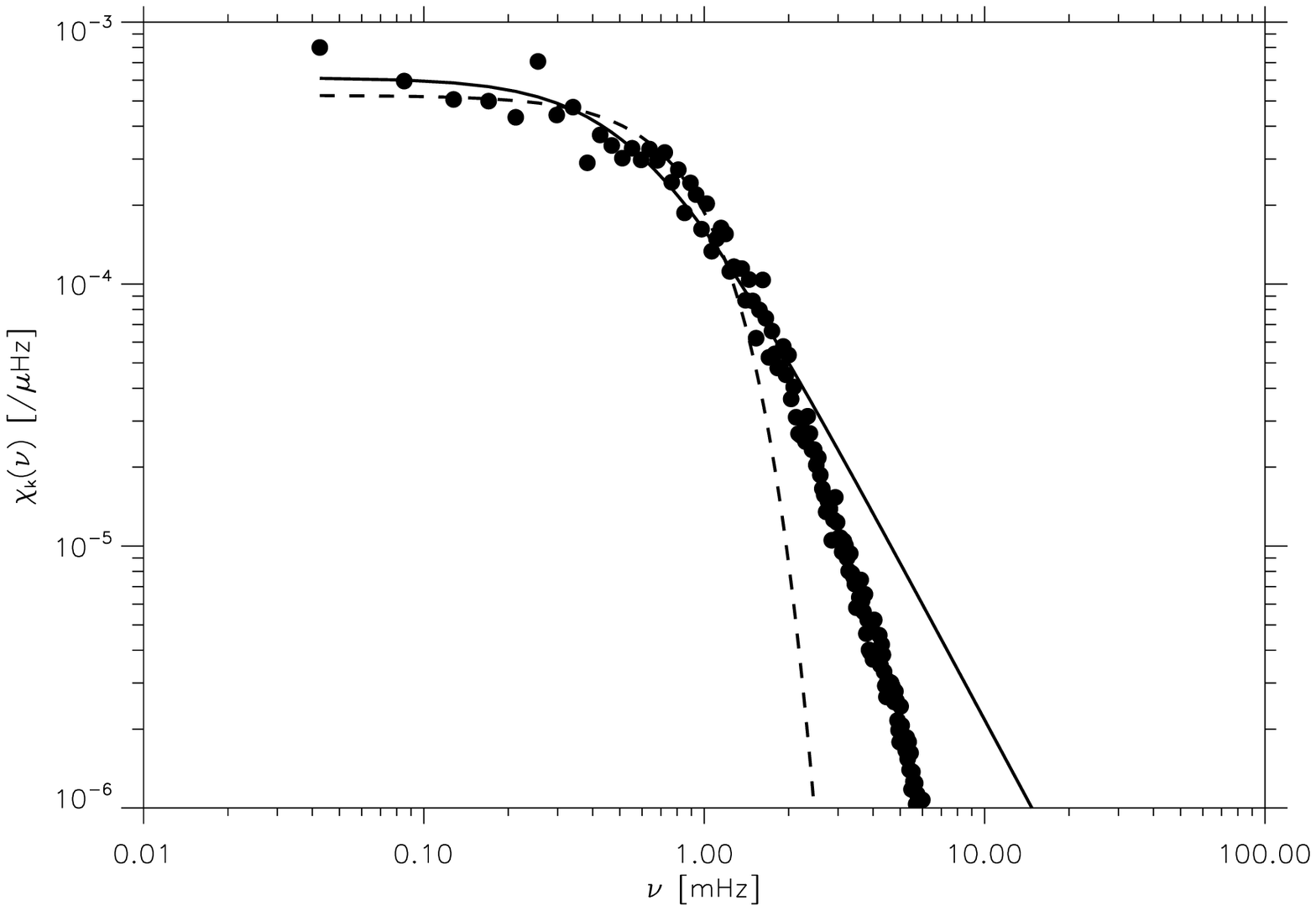}
 \includegraphics[width=10cm]{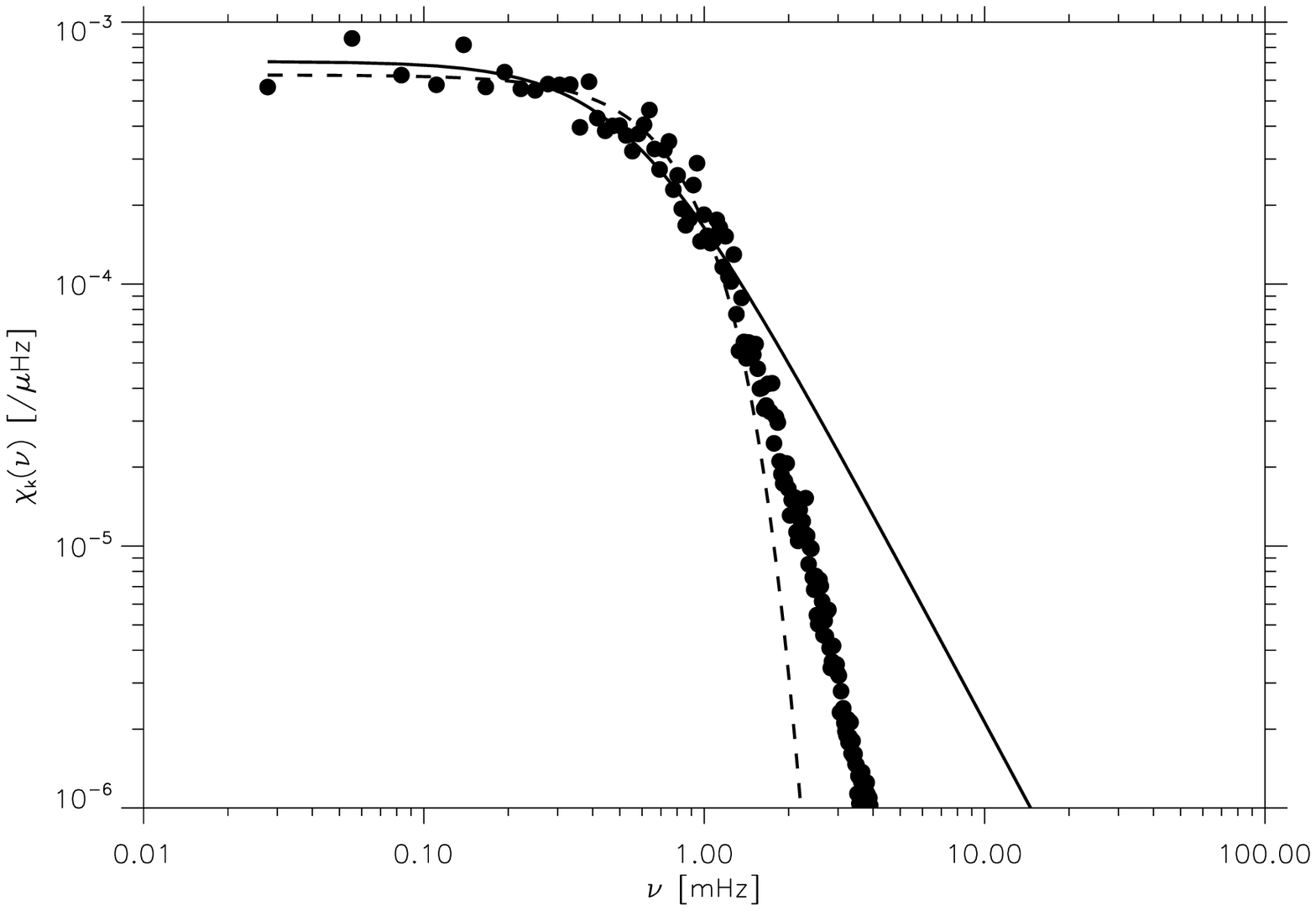} 
\end{center}
 \caption{{\bf Top:} Same as Fig.~\ref{kwpower} for a solar 3D simulation with an horizontal
   resolution of $\simeq$ 50~km \citep{Kevin06a}.
{\bf Bottom:}  Same as top for a solar 3D simulation with an horizontal
   resolution of $\simeq$ 120~km.
}
\label{kwpower_2}
\end{figure}

The excitation of  low-frequency modes ($\nu \lesssim
3$~mHz) is mainly due to the large scale eddies. However, the higher the
frequency the more important the contribution of the small scales.
3D solar simulations show that, at small scales, $\chi_k$ is neither Lorentzian nor
Gaussian \citep{Georgobiani06}.
Hence, according to \citet{Georgobiani06}, it is impossible to separate
the spatial component $E(k)$ from the temporal component at all
scales with the same simple analytical functions.
However, such results are obtained using Large Eddy Simulation
(LES). The way the small scales are
treated in LES can affects our description of turbulence.
Indeed, \cite{He02} have shown that LES results in a
$\chi_k(\omega)$ that decreases at all resolved
scales too rapidly with $\omega$  with respect to direct numerical
simulations (DNS).  Moreover, \citet{Jacoutot08a} found that computed mode
excitation rates  depend significantly  on the adopted sub-grid model.
Furthermore,  \citet{Samadi07a} have shown that, at
a given length scale, $\chi_k$ tends
toward a Gaussian when the spatial resolution is decreased. This is
illustrated in Fig.~\ref{kwpower_2} by comparison with
Fig.~\ref{kwpower}. 
In summary, the numerical resolution or the sub-grid
model can substantially affect our description of the small scales.
Improving the modeling of the excitation of the high frequency modes
then requires  more realistic and more resolved hydrodynamical 3D
simulations.

Up to now, only analytical functions were assumed for $\chi_k(\omega)$.
We have here implemented, for the calculation of $\cal P$,
the eddy time-correlation function derived \emph{directly} from long
time series of 3D simulation realizations with an intermediate
horizontal resolution ($\simeq$ 50~km). 
As shown in
Figs.~\ref{pow_sun_2} and \ref{pow_acena}, the mode excitation rates,
$\cal P$, obtained from $\chi_k^{\rm 3D}$, are  found to be comparable to
that obtained assuming a Lorentzian $\chi_k$, except at
high frequency in the case of the Sun. This is obviously the direct
consequence of the fact that a Lorentzian $\chi_k$ reproduces rather
well $\chi_k^{\rm 3D}$ (see Fig.~\ref{kwpower}), except at high frequency where $\chi_k^{\rm
  3D}$ decreases more rapidly than the Lorentzian function (see
Fig.~\ref{kwpower_2} left). 
At high frequency, calculations based on a Lorentzian $\chi_k$ result
in larger ${\cal P}$ and reproduce better the helioseismic
constraints than those based on $\chi_k^{\rm 3D}$ (see Fig.~\ref{pow_sun_2}). This indicates
perhaps that $\chi_k^{\rm 3D}$ decreases more rapidly with frequency
than it should. This is consistent with \cite{He02}'s results who
found that LES predict a too rapidly decrease with $\nu$ compared to
the DNS (see above). 
  
\begin{figure}
\begin{center}
 \includegraphics[width=10cm]{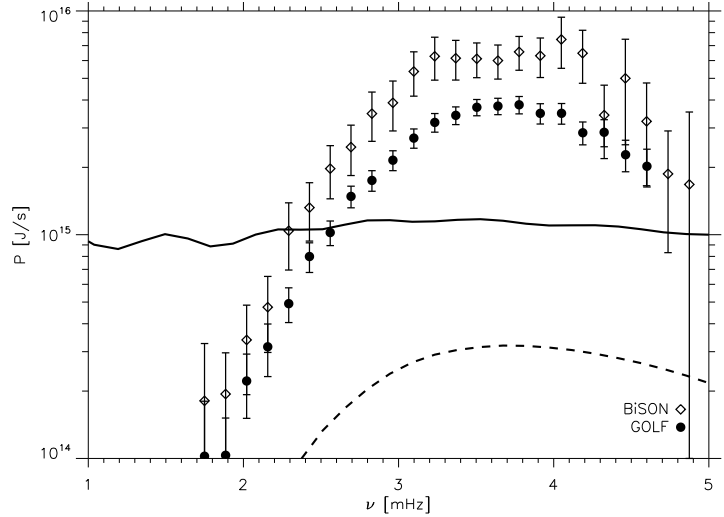} 
\end{center}
 \caption{Same as Fig.~\ref{pow_sun_2}.  The lines correspond to
   calculations using the formalism by \citet{Chaplin05}. Two choices
   for $\chi_k$ was considered : a  Lorentzian $\chi_k$ (solid line)
   and a Gaussian $\chi_k$ (dashed line). In both calculations,
   driving due to the entropy fluctuations is not included. 
}
\label{pow_sun_chaplin05}
\end{figure}

\citet{Chaplin05} also found that the use of a Gaussian
$\chi_k$ severely  under-estimates the observed solar mode excitation rates.
However, in contrast with \citet{Samadi02II}, they mention that a
Lorentzian $\chi_k$ results in a severe over-estimation for the
low-frequency modes. In order to illustrate the results by
\citet{Chaplin05}, we have computed the solar mode
excitation rates using their formalism and a solar envelope
equilibrium model similar to the one considered by these authors
\citep[see][]{Samadi02I}. The result is shown 
in Fig.~\ref{pow_sun_chaplin05}.
We clearly see that the mode excitation rates computed using a Gaussian
$\chi_k$  overestimate by $\sim$ 20 the seismic constraints. 
This  result is consistent with this found  by \citet{Samadi02II}. 
On the other hand, in contrast with \citet{Samadi02II}, the modes with frequency
below $\nu \sim$2\,mHz are severely over-estimated when a  Lorentzian $\chi_k$ is assumed. 
It should be pointed out that the excitation of modes  with
frequency $\nu \lesssim$ 2 mHz occurs in a region more extended than covered by the
solar 3D simulation used by \citet{Samadi02I}. 
On the other hand, the pure 1D modeling by \citet{Chaplin05}, includes 
all of the convective zone.  
The severe over-estimation at low frequency of the mode excitation
rates, is explained by the authors by the fact that, at a given
frequency, a Lorentzian $\chi_k$ decreases too slowly with depth compared to a Gaussian
$\chi_k$. Consequently, for the
low-frequency modes, a substantial fraction of the integrand of \eq{C2R}
arises from large eddies situated deep in the Sun. This might
suggest that, in the deep layers, the eddies that
contribute efficiently have rather a Gaussian $\chi_k$.  
However, this remains an open issue.

\section{Closure models and anisotropy}
\label{closure and anisotropy}

The decomposition of \eq{qna} assumes the quasi-normal approximation
(QNA). However, it is well known that the departure from the QNA is important in a 
strongly turbulent medium. In addition, a closure
model based on the QNA does not ensure the positiveness of the energy
\citep[see details in e.g.][]{Lesieur97}. 
Furthermore, the QNA is strictly valid only for normally distributed
fluctuating quantities with zero mean. However, the upper-most part of the convection zone is a turbulent
convective medium composed of essentially two flows that are
asymmetric with respect to each other. 
Hence, in such a medium, the probability distribution
function of the fluctuations of the vertical velocity and temperature
do not follow a Gaussian law. 
As verified by \citet[B06a hereafter]{Kevin06a} and \citet{Kupka07}, departure from
the QNA is important in the upper part of the solar convective
zone. Indeed,  this approximation   under estimates, in the quasi-adiabatic region,  by
$\approx$~50~\% the fourth-order moment of the vertical velocity derived from 
a solar 3D simulation.  

The term in the LHS of \eq{qna} corresponds to a two-point
correlation product involving the velocity, i.e. $ \langle (u_i \, u_j )_1 \,
(u_k  \, u_l  )_2 \rangle (r,\tau)$ where $r$ and $\tau$ are the
spatial correlation and temporal correlation lengths respectively.
For  $r \rightarrow 0$ and $\tau \rightarrow 0$, this term reduces to
a \emph{one-point} correlation product,  $ \langle u_i \, u_j  \,
u_k  \, u_l  \rangle $, also refered as a fourth-order moment (FOM hereafter). 
As we shall see below, it is possible to derive an
improved closure model for this term that does not rely on the QNA.
However, we still require a  prescription for the \emph{two point}
correlation products involving the velocity ($ \langle (u_i \, u_j )_1 \,
(u_k  \, u_l  )_2 \rangle (r,\tau)$)
 and the entropy fluctuations ($ \langle (\vec u \, s_t )_1 \, (\vec u \, s_t )_2 \rangle (r,\tau)$).
For radial modes or low $\ell$ order modes, only the radial component
of the velocity ($w$) matters. Hence, for these modes we require a
prescription for  $ \langle w^2 _1 \, w^2_2 \rangle (r,\tau)$ and $ \langle (w \, s_t )_1
\, (w \, s_t )_2 \rangle (r,\tau)$. 
By default, \citet{Kevin06b} have proposed that  $ \langle w^2 _1 \, w^2_2 \rangle (r,\tau)$
varies with $r$ and $\tau$ in the same way than in the QNA (\eq{qna}),
that is:
\eqna{
\langle w_1^2 \, w_2^2\rangle  & =  & { {\cal K}_w \over 3} \, \langle w^2_1 \, w^2_2\rangle _{\rm
  QNA} \; ,
\label{2ptcorel}
}
where ${\cal K}_w $ is a constant and $\langle w^2_1 w^2_2\rangle _{\rm QNA}$ is the two-point correlation product
given for $w$ according to the QNA (\eq{qna}). 
Accordingly, the contribution of the Reynolds stress ($C^2_R$, \eq{C2R}) 
is modified as:
\eqna{
C_R^2 & = &  4 \, \pi^{3} \, {\cal G} \, \int_{0}^{M}
dm \, \rho_0 \left (\deriv { \xi_{\rm r}} {r} \right )^2 \, { {\cal K}_w \over 3} S_R(m,\omega_{\rm osc})
\label{C2R_2}
} 
Note that the contribution of the entropy fluctuations ($C^2_S$,
\eq{C2S})  still assumes the QNA. This inconsistency has a small
impact on computed mode excitation rates since $C^2_S$
 is significantly smaller than $C^2_R$, at least for stars that are
 not too hot (but see Sect.~\ref{entropy}).

The constant ${\cal K}_w$ is determined in the limit case where $r \rightarrow 0$
and $\tau \rightarrow 0$. Indeed, when $r \rightarrow 0$ and $\tau
\rightarrow 0$, we have:
\eqna{
\langle w^4 \rangle  & =  &  {  {\cal K}_w \over 3}  \, \langle w^4\rangle _{\rm
  QNA} \; ,
\label{m4}
}
where $ \langle w^4 \rangle $ is by definition the fourth-order moment (FOM
hereafter) associated with $w$ and $\langle w^4\rangle _{\rm   QNA}$ is the one given by the QNA.
In the same way,  \eq{qna} gives :
\eqna{
\langle w^4\rangle _{\rm QNA} & =  & 3 \, \langle w^2\rangle ^2 \; .
\label{m4_qna}
}
Using Eqs.~(\ref{m4}) and (\ref{m4_qna}), we then derive the
constant ${\cal K}_w $:
\eqna{
{\cal K}_w & =& 3 \,  {    {\langle w^4\rangle } \over {\langle w^4\rangle _{\rm QNA}} } =
{ \langle w^4\rangle  \over \langle w^2\rangle ^2} \; ,
\label{kw}
}
which is by definition the Kurtosis. This quantity measures the oblateness of
the probability density function  (see e.g. B06a). For  normally distributed $w$ we have
${\cal K}_w = 3$.  The Kurtosis  then measures  the departure of the FOM from
the QNA.

\begin{figure}
\begin{center}
 \includegraphics[width=10cm]{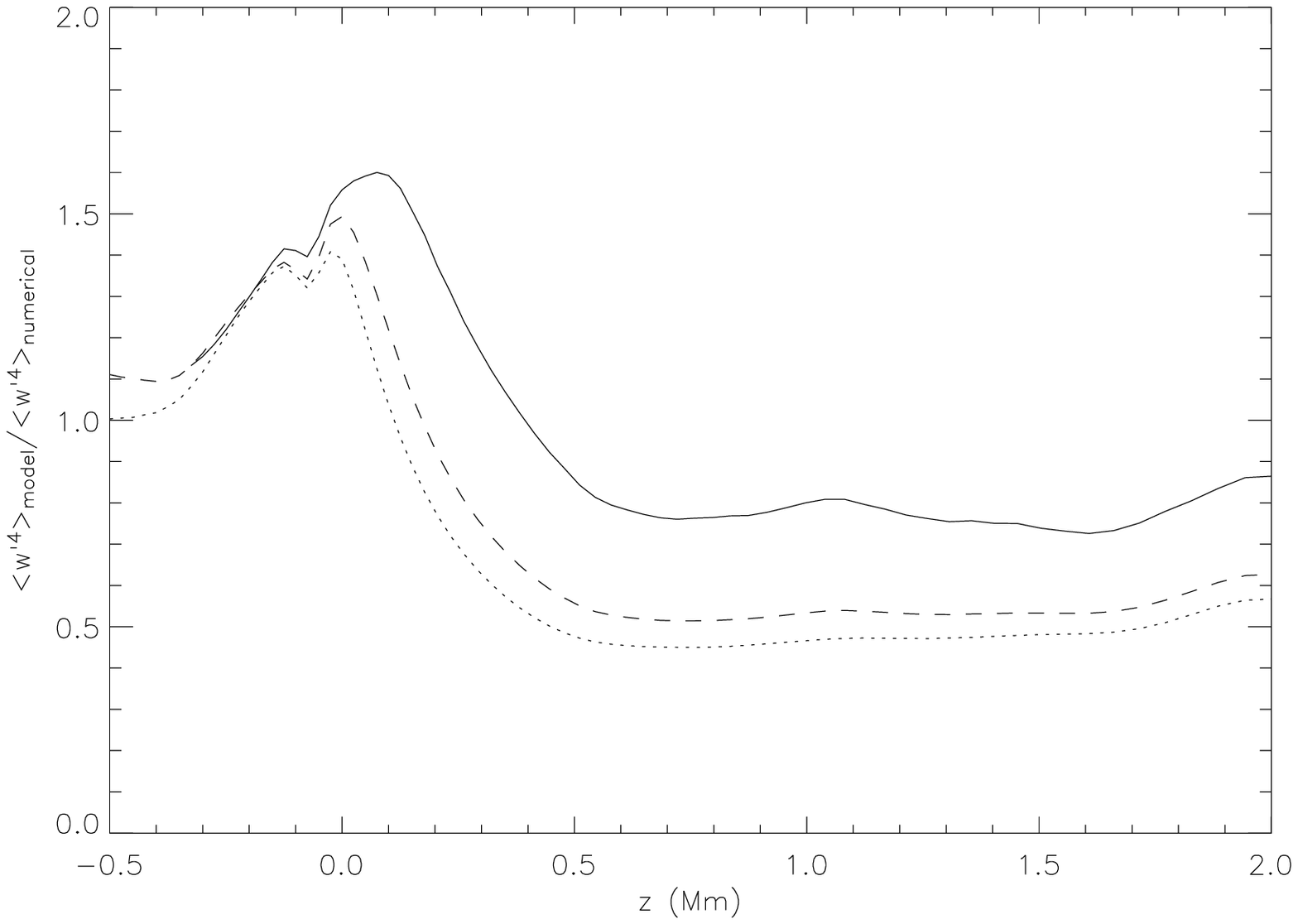}
\end{center}
 \caption{Fourth-order moment (FOM) of the velocity, $\langle w4\rangle  = {\cal
     K}_w \, \langle w2\rangle ^2$, as a
    function of depth $z$, normalized to the FOM derived from the 3D
    simulation.
    In all cases the Kurtosis ${\cal K}_w$, (\eq{kw}) is
    calculated according to \eq{kw_GH} but with different skewness,
    $S_w$.
    The solid line $S_w$ is computed according to B06a's closure
    model, the dashed line assumes \citet{Gryanik02}'s expression for
    $S_w$ (\eq{Sw_GH}) and finally the dotted line assumes the QNA,
    that is $S_w=0$ and ${\cal K}_w=3$.
}
\label{fom}
\end{figure}

Closure models more sophisticated than the QNA can be used. Among
those, the two-scale mass flux model \citep{Abdella97} improved by
\citet{Gryanik02} takes  the asymmetries in the medium  into account
but is only applicable for quasi-laminar 
flows. 
 For ${\cal K}_w$, \citet{Gryanik02} obtained the following
expression:
\eqna{
\label{kw_TFM}
 {\cal K}_w         &=&  (1+S^2_{w})  
}
with  the skewness, $S_w$, given by:
\eqna{
\label{Sw_GH}
  S_{w} & \equiv  & {\langle w^3\rangle  \over \langle w\rangle ^{3/2} } = \frac{1-2a}{\sqrt{a(1-a)}} \; 
}
where $a$ is the mean fractional area occupied by the updrafts in the
horizontal plane. In the QNA limit, i.e. when the random quantities
are distributed according to a Normal distribution with zero mean, we
necessarily have  $S_w=0$. Hence, in the QNA limit, \eq{kw_TFM} does not match
the expected value i.e. ${\cal K}_w=3$. 
Then, \citet{Gryanik02} proposed to modify \eq{kw_TFM} as
follows:
\eqna{
\label{kw_GH}
 {\cal K}_w        &=&  3 \, (1+{1 \over3} \, S_w^2 ) \; .
}
Figure~\ref{fom} shows that the FOM  based on
\eq{kw_GH} with  $S_w$ given by \eq{Sw_GH}, results in a negligible
improvement with respect to the QNA. 
However, when  $S_w$ is derived directly from the 3D simulation and
plugged into \eq{kw_GH}, \eq{kw_GH} is then a very good evaluation
of the FOM derived from a 3D simuation of the outer layer of the Sun as
verified by B06a and \citet[][]{Kupka07}.

\citet{Kevin06a} have  generalized \citet{Gryanik02}'s approach by
taking the skewness introduced by the presence of up- and
down-drafts  \emph{and} the turbulent properties of each flow into
account. Accordingly, they have derived   a more accurate expression for $S_w$
(see the expression in B06a). As shown in Fig.~\ref{fom}, calculations
of the FOM based on \eq{kw_GH} and their expression for $S_w$ reproduce rather
well --~ in the quasi-adiabatic region ~-- the FOM derived from the
solar 3D simulation.

\begin{figure}
\begin{center}
 \includegraphics[width=10cm]{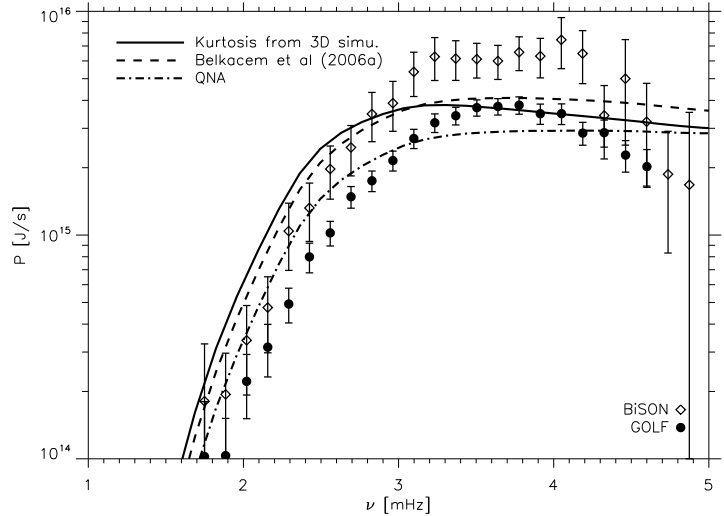} 
\end{center}
 \caption{Same as in Fig.~\ref{pow_sun_2}. The thick lines 
   correspond to calculations where the Reynolds stress contribution
   is computed according to \eq{C2R_2}. 
   The Kurtosis ($K_w$) is computed here in a 
   different manner: for the solid line ${\cal K}_w$  is obtained 
   directly from a 3D solar simulation, for the dashed line the
   Kurtosis is calculated according to \eq{kw_GH} where the skewness
   ($S_w$) is obtained from B06a's closure model, and finally for
   the dot-dashed line we have assumed the QNA, that is $S_w=0$ and  ${\cal K}_w=3$.
}
\label{pow_sun_3}
\end{figure}

\citet{Kevin06b} have computed mode excitation rates,
${\cal P}$ according to \eq{C2R_2} with the Kurtosis ${\cal K}_w$ given
 by \eq{kw_GH} and with the skewness $S_w$  computed according to
 B06a's closure model. The maximum in ${\cal P}$ is found about 30\,\%
 larger than in calculations based on the QNA and fits better the maximum in ${\cal P}$ derived from the
helioseismic data. 
This increase is significantly larger
than the entropy contribution (the term ${\cal S}_S$ in \eq{C2S},
which is of the order of $\sim$~15\,\%, see Sect.~\ref{entropy}). 
We stress that, however, 30\,\% is of the same order as the difference
between seismic constraints of different origins (SOHO/GOLF, GONG, BiSON). These
results are illustrated in Fig.~\ref{pow_sun_3}.

\section{Importance of the stellar stratification and chemical composition}

\subsection{Role of the turbulent pressure}

\citet{Rosenthal99} have shown that taking   the turbulent
pressure into account in a realistic way in the 1D global solar models results in a much better agreement between
observed and theoretical mode frequencies of the Sun.
Following \citet{Rosenthal99}, \citet{Samadi08} have studied the
importance for the calculation of the mode excitation
rates of taking the turbulent pressure into account in the
averaged 1D model. For this purpose, they  have built two 1D models
representative of the star $\alpha$~Cen~A. One model (here refered
as the ``patched'' model), has its surface
layers taken directly from a  fully compressible 
3D hydrodynamical numerical  model.  A second model (here refered
as``standard'' model),  has its surface
layers  computed using standard physics, in particular convection is
described according  \citet{Bohm58}'s  mixing-length local theory of convection (MLT)
and   turbulent pressure is ignored. 

\citet{Samadi08} found that the calculations of ${\cal P}$ involving
eigenfunctions computed on the 
basis of the ``patched'' global 1D model  reproduce  much better  
the seismic data derived for $\alpha$~Cen~A than calculations based on
the eigenfunctions computed with  
the  ``standard'' stellar model, i.e. built with  the 
MLT and ignoring turbulent pressure. 
This is because a model that includes turbulent pressure results in
 \emph{lower} mode masses ${\cal M}$ than a model that ignores turbulent
pressure. 
This can be understood as follows: Within the super-adiabatic region, 
a model that includes turbulent pressure provides an additional
support against gravity,  hence has a lower gas pressure
and density than a model that does not include
turbulent pressure \citep[see also][]{Nordlund99b,Rosenthal99}.
As a consequence,  mode inertia (\eq{inertia}) or equivalently 
mode masses (\eq{MM}) are then \emph{lower} in a
 model that includes turbulent pressure.

\subsection{Role of the surface metal abundance}
\label{role_metal_abundance}

\citet{Samadi09a} have recently studied the role of the surface metal abundance on the
  efficiency of the stochastic driving. 
For this purpose, they have computed two 3D hydrodynamical simulations representative --~ in
effective temperature and gravity ~-- of the
surface layers of HD~49933, a star which is rather metal poor compared
to the Sun since its surface iron-to-hydrogen abundance is
[Fe/H]=-0.37. 
One 3D simulation (hereafter labeled as S0) has a solar
metal abundance  and the other (hereafter labeled as S1) has [Fe/H]
ten times smaller. For each 3D simulation they have build a ``patched'' model  in the manner of
\citet{Samadi08}  and computed the  acoustic modes associated with the  ``patched'' model.

\begin{figure}
\begin{center}
 \includegraphics[width=10cm]{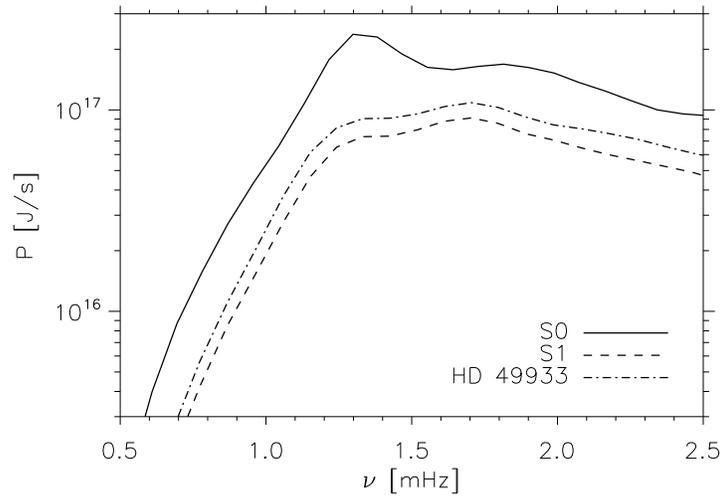}
\end{center}
\caption{Mode excitation rates, $\mathcal P$, as a function
of the mode frequency ($\nu$) obtained for two 3D models with the effective
temperature and the surface gravity of HD~49933 but with two different surface
metal abundances (see Sect.~\ref{role_metal_abundance} and \citet{Samadi09a}). 
The solid line corresponds to the 3D model
with the metal abundance (S0) and the dashed line to metal poor 3D
model (S1). The dot-dashed  line corresponds to the mode excitation
rates derived for the specific case of HD~49933 as explained 
in  \citet{Samadi09a}. 
}
\label{pow_hd49933}
\end{figure}

As seen in Fig.~\ref{pow_hd49933}, the mode excitation rates ${\cal P}$  associated with S1  are found
to be about \emph{three times smaller} than those associated with S0. 
This difference is related to the fact that a lower surface metallicity
results in a lower opacity, and accordingly in an higher
surface density.
In turn, the higher the density, the smaller are the convective
velocities to transport by convection the same amount of
energy. Finally, smaller convective velocities result in a less
efficient driving \citep[for details  see][]{Samadi09a}.
This conclusion is qualitatively consistent with that  by
\citet{Houdek99} who --~ on the basis of a mixing-length
approach ~-- also found that the mode amplitudes decrease with decreasing
metal abundance.

Using the seismic determinations of the mode linewidths measured by
CoRoT for HD~49933 \citep{Benomar09b} and the theoretical
mode excitation rates computed for the specific case of HD~49933, \citet{Samadi09b}
have derived the theoretical mode amplitudes of the acoustic
modes of HD~49933. Except at rather high frequency ($\nu \gtrsim 1.9$~mHz), their amplitude calculations are within approximately 1-$\sigma$ in agreement with the mode amplitudes derived
from the CoRoT data  \citep[for more details see][]{Samadi09b}. They also show
that assuming a solar metal abundance rather than the observed 
metal abundance of the star would result in larger mode amplitudes and
hence in a larger discrepancy with the seismic data.
This illustrates the importance of taking the surface metal abundance of
the solar-like pulsators into account when modeling the mode excitation.

\section{Contribution of the entropy fluctuations}
\label{entropy}

Using the method summarised in Sect.~\ref{direct_calculation},
\citet{Stein01II} have computed \emph{directly} from a 3D simulation of the surface of the
Sun the contribution of the incoherent entropy  fluctuations (\eq{S_S}).
They also found that the entropy fluctuation  is small
compared to the Reynolds stress contribution. 
However, as shown by \citet{Samadi07a}, the relative contribution of the entropy
to the total excitation rate increases rapidly with the effective
temperature, $T_{\rm eff}$. For instance, the solar-like pulsator
HD~49333 has a significantly higher $T_{\rm eff}$ than the
Sun. \citet{Samadi09a}  found that for this star the entropy
fluctuations 
contributes up to $\sim$ 30~\% while it is only about 15~\% in the case
of the Sun \citep[see][]{Samadi07a} and in the case of $\alpha$~Cen~A
\citep[see][]{Samadi08}. 

As pointed-out by \citet{Houdek06}, the solar and stellar 3D
simulations performed by \citet{Stein04} show some  partial canceling between 
the Reynolds stress contribution (${\cal S}_R$, \eq{S_R}) and
contribution due to the entropy (${\cal S}_S$, \eq{S_S}). This
cancellation increases with increasing $T_{\rm eff}$
\citep[see][]{Stein04}. 
In the theoretical model of stochastic excitation, the cross terms
between the entropy fluctuations and the Reynolds 
stresses vanish (see Sect.~\ref{Driving sources}). 
As originally suggested by \citet{Houdek06} and discussed in
\citet{Samadi09b},  the existence of a partial canceling can
decrease the mode amplitude and improve the agreement with the seismic
observations. However, there is currently no theoretical modeling of the
interference between these two terms (see the discussion in
Sect.~\ref{discussion} and in \citet{Samadi09b}. 

\section{Alternative approaches}
\subsection{Energy equipartition}
\label{Energy equipartition}

Under certain conditions that we will emphasize below, GK have shown that 
there is an equipartition of kinetic energy between an acoustic mode and
the resonant eddy.
To derive this principle, GK  assume that the acoustic modes are
damped by turbulent viscosity \emph{and} excited by the Reynolds
stresses. We reproduce here their demonstration. 
For the sake of simplicity, we will consider  modes with 
 $\omega_{\rm osc} \; \tau_0 \lesssim 1$ where $\tau_0$ is
the characteristic time of the energy bearing eddies typically
located in the
upper part of the convective zone, that is the region where the driving is the most
vigorous.   Furthermore, we neglect as did GK the driving by the entropy
fluctuations (\eq{C2S}).
According to Eqs.~(\ref{pow_2}) and (\ref{C2R_propto}), we  have roughly  for acoustic modes with $\omega_{\rm osc} \;
\tau_0 \lesssim 1$:
\eqna{
 {\cal P} & \propto & { 1 \over {I} } \, \int dm \, \left | { {d \xi_{\rm r}}
   \over { dr } } \right |^2  \,  { E_{\rm
    eddy} \,  \Lambda \, u_0  }  \; ,
\label{P_approx}
}
where  $\Lambda$ is the characteristic size  of the energy bearing
eddies,  $u_0$ their characteristic
velocity (\eq{eqn:E:normalisation}),  $\tau_0 = \Lambda
/u_0  $  their characteristic lifetime (\eq{tau_0}), and 
$ E_{\rm     eddy} = (3/2) \, \rho_0 \,  u_0 ^2 \,
\Lambda^3$ their total kinetic energy. 
Let $k_{\rm osc}$ be the vertical oscillation wave number. We have then $d\xi_{\rm r} /
dr  =  i \, k_{\rm osc} \,  \xi_{\rm r}$. We further assume that --~ in the driving region ~-- the acoustic waves are purely
propagating. This assumption then implies $\omega_{\rm osc} = k_{\rm osc} \,
c_s$ where $c_s$ is the sound speed. 
Accordingly, we can simplified \eq{P_approx} as:
\eqna{
 {\cal P} & \propto & { \omega_{\rm osc}^2 \over {I} } \, \int {\rm d} m \, \left ( { \xi_{\rm r}
   \over c_s } \right )^2  \,  { E_{\rm
    eddy} \,  \Lambda \, u_0  }  \; .
\label{P_approx_1}
}

In the region where the mode are excited, $E_{\rm     eddy}$,
$u_0$, and $c_s$ vary quite rapidly. However, again for the sake of
simplicity we will  assume that these quantities are constant and evaluate them at the layer where the excitation is
the most efficient, i.e. at the peak of the super-adiabatic
temperature gradient. 
The integration of \eq{P_approx} can be approximated as
\eqna{
 {\cal P} & \propto & {1 \over I} \, { \left ( \omega_{\rm osc}  \over c_s \right )^2 }  \,  { E_{\rm
    eddy} \,  \Lambda \, u_0  } \,  \int {\rm d} m \, \xi_{\rm r}^2  \;  .
\label{P_approx_2}
}
Using the expression of the mode inertia (\eq{inertia}), 
we  can finally simplify \eq{P_approx_2} as:
\eqna{
 {\cal P} & \propto &  { \left ( \omega_{\rm osc}  \over c_s \right )^2 }  \,  { E_{\rm
    eddy} \,  \Lambda \, u_0  }  \;.
\label{P_approx_3}
}

Modes damped by turbulent viscosity have their damping rates $\eta$
given by \citep{Ledoux58,Goldreich77a},
\eqna{
\eta & \propto & { 1 \over {3 I }}   \, \int {\rm d} m \,
\nu_t \, \left | r \, { d \over {dr}} \, \left (  {\xi_{\rm r} \over r}
\right ) \right |^2
\;,
\label{eta_viscosity}
}
where $\nu_t$ is the turbulent viscosity.
The simplest prescription for $\nu_t$ is the concept of
eddy-viscosity. This consists in assuming $\nu_t = u_0 \, \lambda
= \tau_0 \, u_0 ^2$. Obviously the turbulent medium is
characterized by eddies 
with a large spectrum of size. However, only the eddies for which
$\omega_{\rm osc} \, \tau_\lambda \approx 1$ are expected to
efficiently damp the mode with frequency $\omega_{\rm osc}$. 
 Since we are looking at the modes such that  $\omega_{\rm osc} \;
\tau_\lambda \lesssim 1$,  only the largest eddies efficiently damp the
mode, that is the eddies with size $\Lambda$. Accordingly, we adopt
$\nu_t =  u_0 \, \Lambda$.  With the same simplifications and
assumptions as those used for deriving \eq{P_approx_3}, we can simplify 
\eq{eta_viscosity} as: 
\eqna{
\eta & \propto & \left ( {\omega_{\rm osc} \over  c_s} \right )^2  \, \Lambda  \,
u_0 \;.
\label{eta_viscosity_2}
}
From \eqs{balance_3}, (\ref{P_approx_3}) and
(\ref{eta_viscosity_2}), we then derive the mode kinetic energy:
\eqna{
{ E}_{\rm osc}   & \propto  &  E_{\rm eddy} \; .
\label{equi}
}
\eq{equi} highlights an equipartition of
kinetic energy between an acoustic mode and the resonant eddies. 
\citet{JCD83b} used this ``equipartition principle'' to derive the
first quantitative estimate of solar-like oscillations in stars. 
The relation of \eq{equi} was derived by assuming that modes are damped
by turbulent viscosity. However, as pointed-out by \citet{Osaki90},
theoretical mode line-widths, $\Gamma=\eta/\pi$, computed in the manner of
\citet{Goldreich77a}, { i.e.} assuming a viscous damping,
are underestimated compared to the    
observations. \citet{Gough76} proposed a different prescription
for $\nu_t$. Nevertheless, assuming \citet{Gough76}'s prescription
also results in similar $\Gamma$ \citep[see][]{Balmforth92b}.
On the other hand, \citet{Xiong00} report that the turbulent viscosity
is the dominant source of damping of the radial p modes. 
As discussed recently by \citet{Houdek08}, there is currently no
consensus about the physical processes that contribute dominantly to
the damping of p modes. If the damping due to  turbulent viscosity
turns out to be negligible,  then there is no reason that the balance between the
mode kinetic energy and the kinetic energy of resonant eddies holds in
general.

\subsection{``Direct'' calculation}
\label{direct_calculation}

The model of stochastic excitation presented in Sect.~\ref{Theoretical
  models} is based on several simplifications and assumptions concerning the
turbulence and the source terms. 
There is an alternative approach proposed by  \citet{Stein01I} that does
not rely on such simplifications  and assumptions.
In such approach, the rate at which energy is stochastically injected
into the acoustic modes  is obtained \emph{directly} from 3D
simulations of the outer layers of a star by computing the (incoherent) work performed on the
acoustic mode by turbulent convection.  
In their approach, the  energy input per unit time  into  a given acoustic mode 
is calculated numerically according to Eq.~(74) of \citet{Stein01I}
multiplied by $S$, the area of the simulation box, to get the
excitation rate (in J s$^{-1}$) : 
\begin{equation}\label{dEdt}
{\cal P}_{\rm 3D} (\omega_{\rm osc}) = \frac{\omega_{\rm osc}^2 \, S }{8 \: \Delta \nu \: {\cal E}_{\omega_{\rm osc}}} \,
\left | \int_r {\rm d} r \:  \Delta \hat P_{\rm nad} (r,\omega_{\rm osc})\: { {\partial
\xi_{\rm r}} \over {\partial r} } \right |^2
\, \label{P3D}
\end{equation}
where $\Delta \hat P_{\rm nad}(r,\omega)$ is the discrete Fourier component of the non-adiabatic
pressure fluctuations, $\Delta P_{\rm nad}(r,t)$,  estimated at the mode
eigenfrequency  
$\omega_{\rm osc}= 2 \pi \nu_0$, $\xi_{\rm r}$ is the radial component
of the  mode displacement eigenfunction,  $\Delta \nu=1/T_s$
the frequency resolution corresponding to the total simulation time
$T_s$ and  ${\ds {\cal E}_{\omega_{\rm osc}}}$  is the normalised mode energy per unit
surface area defined  in \citet[][ their Eq.~(63)]{Stein01I} as:
\begin{equation}
{\cal E}_{\omega_{\rm osc}}= { 1 \over 2 } \, \omega_{\rm osc}^2 \, \int_r dr \: \xi_{\rm r}^2 \, \rho \left ( { r \over R } \right )^2 \; .
\end{equation}

Eq.\ (\ref{dEdt}) corresponds to the calculation of the $P
dV$ work associated with the non-adiabatic gas and turbulent pressure (Reynolds stress) fluctuations.    
In contrast to the pure theoretical models (see Sect.~\ref{Theoretical models}), the
derivation of \eq{dEdt} does not  rely on a simplified model of turbulence. 
For instance,  the relation of \eq{phi_ij} is no longer required.
Furthermore, they do not assume that entropy fluctuations behave as a
passive scalar (\eq{scalaire_passif}). 
However, as for the theoretical models, it is assumed that $\xi_{\rm r}$ varies
on a scale-length larger than the eddies that contributes effectively
to the driving (this is the so-called ``length-scale separation'', see
Sect.~\ref{Length scale separation}). In addition, \eq{dEdt} implicitly assumes the
quasi-Normal approximation (\eq{qna}). 

The expression of \eq{P3D} has been applied to the case of the Sun by
\citet{Stein01II}. These authors obtain a rather good agreement
between $ {\cal P}_{\rm 3D}$ (\eq{P3D}) and the solar mode excitation
rates derived from the GOLF instrument by
\citet{Roca_Cortes99}. However, solar mode excitation rates derived by
\citet{Stein01II} from the seismic analysis by \citet{Roca_Cortes99}
are --~ for a reason that remains to be understood~-- systematically
lower than those derived from the  seismic analysis by
\citet{Baudin05}. 
\citet{Stein04} have computed ${\cal P}_{\rm 3D}$ (Eq.~\ref{P3D}) for
a set of stars located near the main sequence from K to F and a
subgiant K IV star. The comparison between these calculations and
those based on SG's formalism has been undertaken by
\citet{Samadi07a}.  The maximum in ${\cal P}_{\rm 3D}$ was found 
systematically lower than those from calculations based on SG's formalism
(Eqs.~(\ref{pow_2})-(\ref{SS})). These  systematic differences were
attributed by \citet{Samadi07a} to the low spatial resolution of
the hydrodynamical 3D simulations computed by \citet{Stein04}.

\section{Stochastic excitation across the HR diagram}

\subsection{Mode excitation rates}

Using several 3D simulations of the surface of main sequence stars,
\citet{Samadi07a} have shown that the maximum of the mode excitation 
rates, ${\cal P}_{\rm max}$, varies with the ratio $L/M $ as
$\displaystyle{ (L/M)^\alpha}$ where $L$ and $M$ are the luminosity and the
mass of the star respectively and $\alpha$ is the slope of this scaling law.  
Furthermore, they found that the slope $\alpha$ is rather sensitive to the
adopted function for $\chi_k$:  $\alpha$=3.1 for a
Gaussian $\chi_k$ and $\alpha$=2.6 for a Lorentzian one.

The increase of  ${\cal P}_{\rm max}$ with $L/M$ is not
 surprising: It should first be noticed that, even though the ratio $L/M$ is the ratio of two
 global stellar quantities, it nevertheless essentially characterizes 
 the properties of the  stellar surface layers where the mode
 excitation is located since  $L/M \propto T_{\rm eff}^4/g$.  Indeed,
 by definition of the effective temperature, $T_{\rm eff}$, and the stellar radius $R$, the
 total luminosity of the star, $L$, is  given by the Steffan's law: $L = 4 \pi \sigma T_{\rm
   eff}^4 \, R^2$ where $\sigma$ is Steffan's constant.
Furthermore, the surface gravity is $g= GM /R^2$ where $G$ is the
gravitational constant. Accordingly,  $L/M \propto T_{\rm eff}^4/g$. 

Second, as we will show now, it is possible to roughly explain the
dependence of  ${\cal P}_{\rm max}$ with $g$ and $T_{\rm eff}^4$. 
\eq{P_approx_3} can  be rewritten as:
\eqna{
{\cal P} & \propto & { \left ( \omega_{\rm osc}  \over c_s \right )^2 }  \,  { F_{\rm
    kin} \,  \Lambda^4  }  \; .
\label{P_approx_4}
}
 where  
\eqn{
F_{\rm kin}= {3 \over 2}\, \rho_0 \, u_0^3
\label{F_kin}
}  is by  definition the flux of kinetic energy per unit volume\footnote{for the sake of
simplicity we assume here an isotropic medium, accordingly the  flux of
kinetic energy is the same in any direction} and $u_0$ is the characteristic
velocity given by \eq{eqn:E:normalisation}.

The characteristic size $\Lambda$ is approximately
proportional to the pressure scale height $H_p$
\citep[see e.g.][]{Samadi08}. 
From hydrostatic equilibrium, we have $P = \rho \, g \, H_p$. 
Assuming now the equation of state of a perfect gas, we then derive
$H_p \propto T/g$. The sound  speed is given by the relation $c_s^2
= \Gamma_1 \, P /\rho$. Accordingly, using again the perfect gas equation,  we then have $c_S^2 \propto T$. 
From these simplifications, we can simplify \eq{P_approx_4} as:
\eqna{
{\cal P} & \propto &  \omega_{\rm osc}^2  \,   F_{\rm
    kin} \,  { {T^3} \,  g^{-4} }   \; .
\label{P_approx_4_2}
}
 
In the framework of the mixing-length approach, it can be shown that
$F_{\rm kin}$ is roughly proportional to the convective flux $F_c$. 
Indeed, in this framework, the eddies are accelerated by the buoyancy
force over a distance equal to the mixing-length $\Lambda= \alpha \,
H_p$ where $\alpha$ is the mixing-length parameter.
Accordingly, the kinetic energy of the eddies, $ E_{\rm eddy}$,  is
given by \citep[see the lecture notes by][]{Bohm89}
\eqn{
E_{\rm eddy} \equiv {3 \over 2} \, \rho \, u_0^2 \, \Lambda^3 = g \, ( \Delta
\rho \Lambda^3) \, \Lambda
\label{E_eddy}
}
where $\Delta \rho$ is the difference between the density of the eddy
and its surroundings. 
In the Boussinesq approximation, the perturbation of the equation of
state gives:
\eqn{ 
\frac {\Delta \rho }{ \rho} \propto  \frac {\Delta T }{ T}
\label{Delta_rho}
}
where $\Delta T$ is the difference between the temperature of the eddy
and its surrounding.
Now, the convective flux (also referred to as the enthalpy flux) is by
definition the quantity:
\eqn{
F_c \equiv u_0 \, \left ( \rho \, C_p \, \Delta T  \right ) 
\label{F_c}
}
where $c_p =  (\partial s / \partial \ln T)_p$.
Finally, from the definition of \eq{F_kin}  and the set of Eqs.
(\ref{E_eddy})-(\ref{F_c}), one derives $F_{\rm kin} \propto { {g \,
    \Lambda} / T }  \, F_{c} $ and, since $\Lambda \propto T/g$, we show
finally that $ F_{\rm kin} \propto  F_{c} $.

In the region where the driving is the most efficient,
the total energy flux, $F_{\rm tot}$,  is no longer totally
transported by convection (that is $F_c \langle  F_{\rm tot}$). 
However, in order to derive an expression that depends only on the
surface parameters of the star, we will assume that all of the  energy is transported by
convection ;  that is  $ F_c \approx F_{\rm tot} = \sigma \, T_{\rm eff}^4 \propto g
\, (L/M) $ where $\sigma$ is the Steffan's constant.
Accordingly, \eq{P_approx_4_2} can be further simplified as:
\eqna{
{\cal P} & \propto &  \omega_{\rm osc}^2  \,  {
  { T_{\rm eff}^4 \, T^3  } \, g^{-4} } \approx \omega_{\rm osc}^2  \,  {
  { T_{\rm eff}^7  } \,  g^{-4} }  \;,
\label{P_approx_5}
}
where we have assumed $T = T_{\rm eff}$.

Let now defines $\nu_{\rm max}= \omega_{\rm osc}^{\rm max}/2 \pi$ the peak frequency associated with ${\cal
  P}$.  This characteristic frequency can be estimated according to:
\eqn{
\nu_{\rm max} \approx {u_0 / \Lambda} 
}
where the quantity $u_0 / \Lambda$ is estimated in the layer
  where $u_0$ is maximum. 
Using similar simplifications as used previously for ${\cal P}$, we
can show that
\eqn{
\nu_{\rm max}
\propto g \, (T_{\rm eff} / \bar \rho )^{1/3} \;,
\label{nu_max}
} where $\bar \rho$ is
the mean density at the photosphere. We assume that $\bar \rho$ is equal to the
star mean density, that is $\bar \rho \approx M/R^3 \propto g / R$. Accordingly, we then derive from \eq{P_approx_5}
and \eq{nu_max}:
\eqna{
{\cal P}_{\rm max} & \propto & \left ( T_{\rm eff}^4 \right )^{23/12}
\,  g^{-3} \, M^{1/3} \; ,
\label{Pmax_approx}
}
where $M$ is the stellar mass.
For main sequence stars lying in the domain where solar-like
oscillations are
expected, $M^{1/3}$ varies very slowly such that it can be ignored in \eq{Pmax_approx}. 
Then, \eq{Pmax_approx} can finally be  simplified as:
\eqna{
{\cal P}_{\rm max} & \propto & \left ( T_{\rm eff}^4 \right )^{2}
\,  g ^{-3} \; .
\label{Pmax_approx_2}
}
We now clearly see from \eq{Pmax_approx_2} that  ${\cal
  P}_{\rm max}$ as expected
increases with increasing $F_{\rm tot} = \sigma \, T_{\rm eff}^4$ and
decreases with increasing $g$. 

\subsection{Mode surface velocity}

Prior to the CoRoT mission, only crude and indirect derivations of the
averaged mode linewidth had been 
proposed for a few stars \citep[see][]{Kjeldsen05,Fletcher06,Kjeldsen08}.  However, for the
majority of solar-like pulsators observed so far from the ground in
Doppler velocity, such measurements are
not available,  only the maximum of the mode surface
velocity ($V_{\rm max}$ hereafter) is in general accessible. 
For the numerous solar-like pulsators observed from the ground,  we must compute the mode surface
velocity according to \eq{v_s}, which requires the knowledge of not only ${\cal P}$ but
also of the mode damping rates ($\eta=\pi\,\Gamma$).

\citet{Houdek99} have computed $\eta$ for a large set of main sequence models. Using
\citet{Balmforth92c}'s formulation of stochastic excitation, 
 they have also computed the mode excitation rates (${\cal
  P}$).  From their theoretical computations of ${\cal P}$ and $\Gamma=\eta/\pi$,
 they have  derived $v_s$ according to \eq{v_s}. 
Their theoretical calculations for $V_{\rm max}$ result in a scaling
law of the form $(L/M)^\beta$ with a exponent $\beta$=1.5 \citep[see][]{Houdek99}. 

\begin{figure}
\begin{center} \includegraphics[width=10cm]{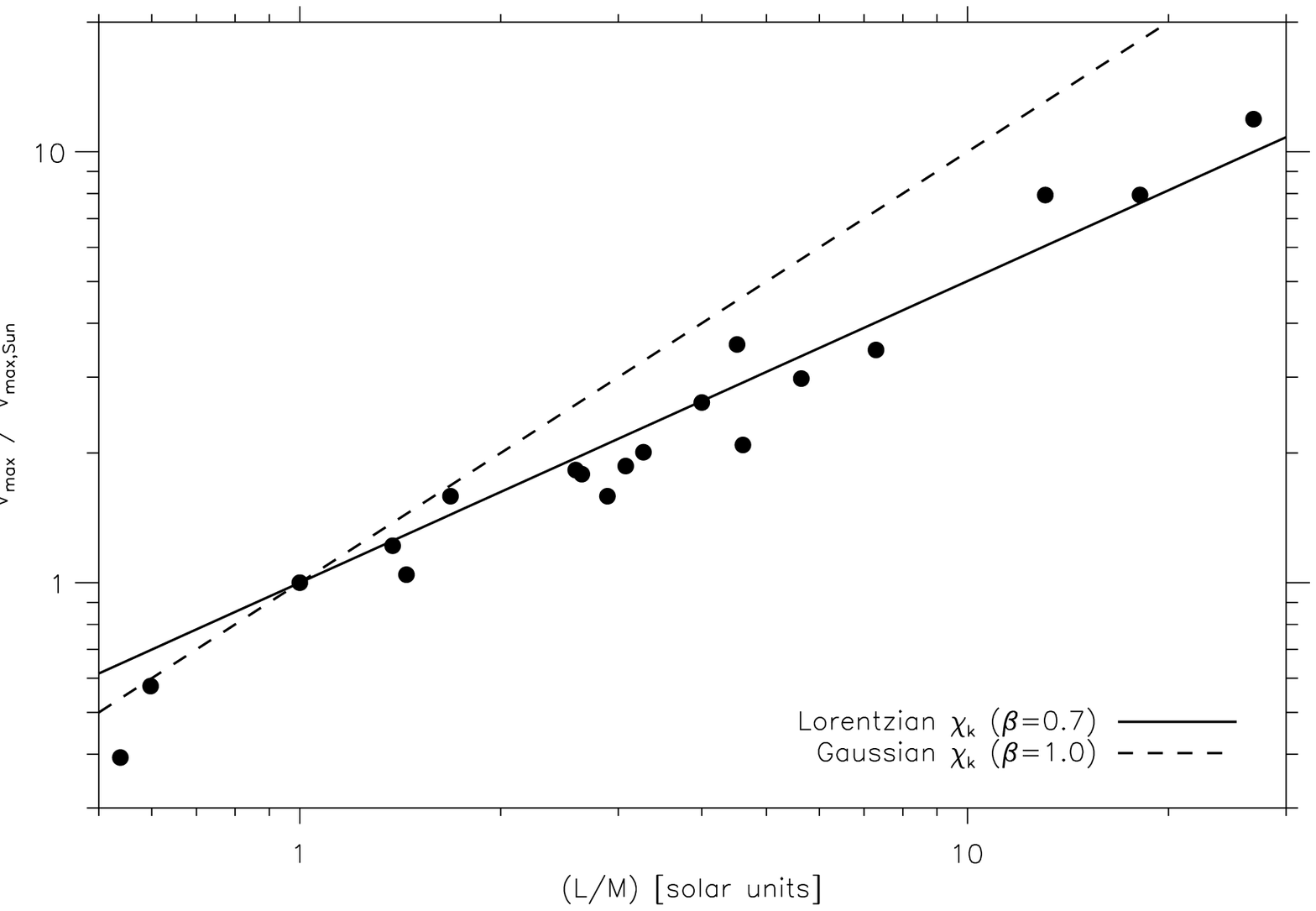}  \end{center}
 \caption{Ratio between $V_{\rm max}$ the maximum of the mode velocity
  relative to the observed solar value ($V_{\rm max}^\odot$= 25.2 cm/s for
  $\ell=1$ modes, see \citealt{Kjeldsen08}). Filled dots
  correspond to the stars for which solar-like oscillations have been
  detected in Doppler velocity \citep[see a detailed list of references
    in][]{Bedding07}. The lines --~ except the dot-dashed line ~--  correspond to the power laws 
obtained from the predicted scaling laws for  ${\cal P}_{\rm max}$ and estimated values of the 
damping rates $\eta_{\rm max}$ (see text for details).
Results for two different eddy time-correlation functions, $\chi_k$,
are presented:  Lorentzian $\chi_k$ (solid line) and Gaussian $\chi_k$
(dashed line).  
}
\label{vmax}
\end{figure}

We have plotted in Fig.~\ref{vmax} the quantity $V_{\rm max}$
associated with the solar-like pulsators  observed so far in Doppler velocity.  Clearly,
$V_{\rm max}$ increases as $(L/M)^{\beta}$ where the exponent $\beta
\simeq 0.7$.  A similar scaling law with the exponent
$\beta=1$ was earlier derived by \citet{Kjeldsen95} from the theoretical
calculations by \citet{JCD83b}. 
 \citet{Houdek99}'s scaling law significantly over-estimates the
mode amplitudes  in F-type stars. For instance for Procyon ($T_{\rm eff} \simeq $  6480~K, $L
\simeq 6.9~L_\odot$ and $L/M \simeq$~4.6), this scaling law over-estimates  $V_{\rm max}$
by a factor $\sim 4$.  

\citet{Samadi07a} have derived $V_{\rm max}$ using mode
damping rates computed by  \citet{Houdek99} and the different scaling
laws found for ${\cal P}_{\rm max} \propto \displaystyle{
  (L/M)^\alpha}$. 
They also found that $V_{\rm max}$ scales as $(L/M)^{\beta}$. This is
not surprising since ${\cal P}_{\rm max}$ varies as
$\displaystyle{ (L/M)^\alpha}$.
Furthermore, the exponent $\beta$ is found to depend significantly on the
choice of $\chi_k$:  $\beta=0.7$ for a Lorentzian $\chi_k$ and 
 $\beta=1$ for a Gaussian  $\chi_k$. As shown in Fig~\ref{vmax},
the best agreement with the observations is found when a Lorentzian
$\chi_k$ is assumed. On the other hand, assuming a Gaussian  $\chi_k$
results in a larger exponent $\beta$. When theoretical mode amplitudes are
calibrated with respect to the solar mode amplitudes, calculations based on a  Gaussian
$\chi_k$  over-estimate the amplitudes of solar-like pulsators significantly
more luminous than the Sun. 

Theoretical calculations by \citet{Houdek99} assume a Gaussian
$\chi_k$. Then according to \citet{Samadi07a}'s results, the too large
value found for $\beta$ by \citet{Houdek99} can partially be explained by the use
of a Gaussian $\chi_k$. 
However, according to \citet{Houdek06}, their too high value
of $\beta$  might  be explained essentially by the  mode damping rates
that could be under-estimated  by a factor $\sim 1.8$.

\section{Discussion and perspectives}
\label{discussion}

The way mode excitation by turbulent convection is modeled is still
very simplified. As discussed below, several approximations must be improved, some assumptions or
hypothesis must be removed.

As seen in Sect.~\ref{Eddy time-correlation}, the driving efficiency crucially depends on the eddy
time-correlation ($\chi_k$). Current models assume that $\chi_k$ varies
with $\omega$ in the same way at any  length scale. 
At the length scale  of the energy bearing eddy, there are some strong indications
that  $\chi_k$ is Lorentzian rather than Gaussian. However, at smaller
scale, it is not yet clear what is the correct description for
$\chi_k$.
Use of more realistic 3D simulations would be very helpful to represent
the correct dynamic behavior of the small-scales.

Current theoretical models that include the entropy fluctuations in
the driving assume that the entropy fluctuations behave as a passive
scalar (see Sect.~\ref{Driving sources}). As a consequence, cross terms between
${\cal S}_R$ and ${\cal S}_S$ vanish.
This is a \emph{strong hypothesis} that is unlikely to be valid in the
super-adiabatic part of the convective zone where driving by the
entropy is important. Indeed, the
super-adiabatic layer is a place where the radiative losses of the
eddies are important because of the optically thin layers. Assuming
that the entropy (or equivalently the 
temperature) is diffusive (\eq{scalaire_passif}) is no longer valid. 
Furthermore, departure from incompressible turbulence is the
largest in that  layer  and, accordingly, 
the cross terms between ${\cal S}_R$ and ${\cal S}_S$ no longer
vanish (see SG).   
Therefore, the passive scalar assumption  is not 
valid in the super-adiabatic layers. 
To avoid this  assumption, one needs to include  the radiative losses in the
modeling. 

One other approximation concerns the spatial separation between the
modes and the contributing eddies. This approximation is less valid in
the super-adiabatic region where the 
turbulent Mach number is no longer small, in particular for high
$\ell$ order modes.
This spatial separation can however be avoided if the kinetic energy
spectrum associated with the turbulent elements ($E(k)$) is properly coupled
with the spatial dependence of the modes (work in progress).

\bigskip

The CoRoT mission, launched 27  December, is  precise enough to
detect solar-like oscillations with amplitudes as low as the solar p
modes \citep{Michel08}.  
Furthermore, thanks to its long term (up to 150~days) and \emph{continuous}
observations, it is possible with CoRoT to resolve solar-like
oscillations, and hence to measure not only the mode amplitudes but
also \emph{directly} the mode linewidths \citep[see
  e.g.][]{Appourchaux08}.  Similarly as in the case of the
Sun, it is now possible with CoRoT to derive direct
constraints on ${\cal P}$ for stars with different characteristics: evolutionary status,
effective temperature, gravity, chemical composition, magnetic field,
rotation, surface convection, ... etc.
We emphasize below some physical processes and conditions that we expect to
 address thanks to the CoRoT data.

Some solar-like pulsators are young stars that show rather strong
activity (e.g. HD~49933, HD~181420,HD~175726, HD~181906,  ...).  
A high level of activity is often linked to the  presence of strong magnetic field.  
Effects of the magnetic field are not taken into account in the
calculation of the mode excitation rates.  
A strong magnetic field can more or less inhibit convective transport 
\citep[see e.g.][]{Proctor82,Vogler05}. Furthermore, as shown by
\citet{Jacoutot08b}, a strong magnetic field can significantly change the way
turbulent kinetic energy is spatially distributed and leads to a less
efficient driving of the acoustic modes. 
In that framework, the CoRoT target HD~175726 is probably an
interesting case. Indeed,  this star shows both a particularly high level of activity and 
solar-like oscillations with amplitudes significantly lower than
expected \citep{Mosser09}.

Young and active stars rotate usually faster than the Sun. As shown
recently by \citet{Kevin09b}, the presence of rotation introduces
additional sources of driving. However, in the case of a moderate rotator such as
HD~49933, these additional sources of driving remain negligible
compared to the Reynolds stress and the entropy source term. 
On the other hand, the presence of rotation has an indirect effect on mode
driving through the modification of the mode eigenfunctions.  
An open issue is: will the CoRoT or the Kepler mission be able to
test the expected effect of rotation \citep[see ][]{Kevin09b}?

Solar-like oscillations have now been firmly detected in several
red giant stars, from both Doppler velocity measurements \citep[see
  the review by ][]{Bedding06} as well as from space based photometry
measurements \citep{Barban07,deRidder06}. More recently, detection of
solar-like oscillations by CoRoT in a huge number of red giant stars
has been announced by \citet{deRidder09}. 
Why look at solar-like oscillations in red giant stars? 
Toward the end of their lives, stars like the Sun greatly expand to
become giant stars. A consequence of this great expand, is the existence of
a very dilute convective envelope. A low density favors a vigorous convection,
hence higher  Mach numbers ($M_t$).
The theoretical models of stochastic excitation  are
strictly valid in a medium where $M_t$ is --~ as
in the Sun and $\alpha$~Cen~A ~--  rather small. Hence, the higher $M_t$, the more
questionable the different approximations and the assumptions involved in the
theory. Hence, red giant stars allow us to test the theory
of mode driving by turbulent in more extreme conditions.

Finally, most of theories of stochastic excitation are developed for radial
modes only. \citet{Dolginov84}, GMK and \citet{Kevin08} have considered the non-radial
case.  
There are interesting applications of such non-radial formalisms, 
for instance the case of solar g modes \citep{Kevin09}, but also
g modes in massive stars that can in principle be excited in
their central convective zones \citep{Samadi09c}. 

\section*{Acknowledgment}
I am very grateful to Marie-Jo Goupil and K\'evin Belkacem for their
valuable comments and advise. I am indebted to J. Leibacher for his
careful reading of the manuscript. I am grateful to the organizers of
the CNRS school of St-Flour for their invitation and I thank the CNRS
for the financial support. 

\bibliographystyle{aa}

\begin{thebibliography}{79}
\expandafter\ifx\csname natexlab\endcsname\relax\def\natexlab#1{#1}\fi

\bibitem[{{Abdella} \& {McFarlane}(1997)}]{Abdella97}
{Abdella}, K. \& {McFarlane}. 1997, J. Atm Phys, 54, 1850

\bibitem[{{Appourchaux} {et~al.}(2008){Appourchaux}, {Michel}, {Auvergne},
  {Baglin}, {Toutain}, {Baudin}, {Benomar}, {Chaplin}, {Deheuvels}, {Samadi},
  {Verner}, {Boumier}, {Garc{\'{\i}}a}, {Mosser}, {Hulot}, {Ballot}, {Barban},
  {Elsworth}, {Jim{\'e}nez-Reyes}, {Kjeldsen}, {R{\'e}gulo}, \&
  {Roxburgh}}]{Appourchaux08}
{Appourchaux}, T., {Michel}, E., {Auvergne}, M., {et~al.} 2008, \aap, 488, 705

\bibitem[{{Balmforth}(1992{\natexlab{a}})}]{Balmforth92c}
{Balmforth}, N.~J. 1992{\natexlab{a}}, \mnras, 255, 639

\bibitem[{{Balmforth}(1992{\natexlab{b}})}]{Balmforth92b}
{Balmforth}, N.~J. 1992{\natexlab{b}}, \mnras, 255, 632

\bibitem[{{Barban} {et~al.}(2007){Barban}, {Matthews}, {de Ridder}, {Baudin},
  {Kuschnig}, {Mazumdar}, {Samadi}, {Guenther}, {Moffat}, {Rucinski},
  {Sasselov}, {Walker}, \& {Weiss}}]{Barban07}
{Barban}, C., {Matthews}, J.~M., {de Ridder}, J., {et~al.} 2007, \aap, 468,
  1033

\bibitem[{Batchelor(1970)}]{Batchelor70}
Batchelor, G.~K. 1970, The theory of homogeneous turbulence (University Press)

\bibitem[{{Baudin} {et~al.}(2005){Baudin}, {Samadi}, {Goupil}, {Appourchaux},
  {Barban}, {Boumier}, {Chaplin}, \& {Gouttebroze}}]{Baudin05}
{Baudin}, F., {Samadi}, R., {Goupil}, M.-J., {et~al.} 2005, \aap, 433, 349

\bibitem[{{Bedding} \& {Kjeldsen}(2006)}]{Bedding06}
{Bedding}, T.~R. \& {Kjeldsen}, H. 2006, Memorie della Societa Astronomica
  Italiana, 77, 384

\bibitem[{{Bedding} \& {Kjeldsen}(2007)}]{Bedding07}
{Bedding}, T.~R. \& {Kjeldsen}, H. 2007, Communications in Asteroseismology,
  150, 106

\bibitem[{{Belkacem} {et~al.}(2009{\natexlab{a}}){Belkacem}, {Mathis},
  {Goupil}, \& {Samadi}}]{Kevin09b}
{Belkacem}, K., {Mathis}, S., {Goupil}, M.~J., \& {Samadi}, R.
  2009{\natexlab{a}}, in press (astro-ph/0909.148)

\bibitem[{{Belkacem} {et~al.}(2008){Belkacem}, {Samadi}, {Goupil}, \&
  {Dupret}}]{Kevin08}
{Belkacem}, K., {Samadi}, R., {Goupil}, M.-J., \& {Dupret}, M.-A. 2008, \aap,
  478, 163

\bibitem[{{Belkacem} {et~al.}(2009{\natexlab{b}}){Belkacem}, {Samadi},
  {Goupil}, {Dupret}, {Brun}, \& {Baudin}}]{Kevin09}
{Belkacem}, K., {Samadi}, R., {Goupil}, M.~J., {et~al.} 2009{\natexlab{b}},
  \aap, 494, 191

\bibitem[{{Belkacem} {et~al.}(2006{\natexlab{a}}){Belkacem}, {Samadi},
  {Goupil}, \& {Kupka}}]{Kevin06a}
{Belkacem}, K., {Samadi}, R., {Goupil}, M.~J., \& {Kupka}, F.
  2006{\natexlab{a}}, \aap, 460, 173

\bibitem[{{Belkacem} {et~al.}(2006{\natexlab{b}}){Belkacem}, {Samadi},
  {Goupil}, {Kupka}, \& {Baudin}}]{Kevin06b}
{Belkacem}, K., {Samadi}, R., {Goupil}, M.~J., {Kupka}, F., \& {Baudin}, F.
  2006{\natexlab{b}}, \aap, 460, 183

\bibitem[{{Benomar} {et~al.}(2009){Benomar}, {Baudin}, {Campante}, {Chaplin},
  {Garc{\'{\i}}a}, {Gaulme}, {Toutain}, {Verner}, {Appourchaux}, {Ballot},
  {Barban}, {Elsworth}, {Mathur}, {Mosser}, {R{\'e}gulo}, {Roxburgh},
  {Auvergne}, {Baglin}, {Catala}, {Michel}, \& {Samadi}}]{Benomar09b}
{Benomar}, O., {Baudin}, F., {Campante}, T.~L., {et~al.} 2009, \aap, 507, L13

\bibitem[{{B\"ohm-Vitense}(1958)}]{Bohm58}
{B\"ohm-Vitense}, E. 1958, Zeitschr. Astrophys., 46, 108

\bibitem[{{Bohm-Vitense}(1989)}]{Bohm89}
{Bohm-Vitense}, E. 1989, Introduction to stellar astrophysics, Vol.~3
  (Cambridge University Press)

\bibitem[{{Chaplin} \& {Basu}(2008)}]{Chaplin08}
{Chaplin}, W.~J. \& {Basu}, S. 2008, \solphys, 36

\bibitem[{{Chaplin} {et~al.}(1998){Chaplin}, {Elsworth}, {Isaak}, {Lines},
  {McLeod}, {Miller}, \& {New}}]{Chaplin98}
{Chaplin}, W.~J., {Elsworth}, Y., {Isaak}, G.~R., {et~al.} 1998, \mnras, 298,
  L7

\bibitem[{{Chaplin} {et~al.}(2005){Chaplin}, {Houdek}, {Elsworth}, {Gough},
  {Isaak}, \& {New}}]{Chaplin05}
{Chaplin}, W.~J., {Houdek}, G., {Elsworth}, Y., {et~al.} 2005, \mnras, 360, 859

\bibitem[{{Christensen-Dalsgaard} \& {Frandsen}(1983)}]{JCD83b}
{Christensen-Dalsgaard}, J. \& {Frandsen}, S. 1983, Solar Physics, 82, 469

\bibitem[{{Cowling}(1941)}]{Cowling41}
{Cowling}, T.~G. 1941, \mnras, 101, 367

\bibitem[{{de Ridder} {et~al.}(2009){de Ridder}, {Barban}, {Baudin}, {Carrier},
  {Hatzes}, {Hekker}, {Kallinger}, {Weiss}, {Baglin}, {Auvergne}, {Samadi},
  {Barge}, \& {Deleuil}}]{deRidder09}
{de Ridder}, J., {Barban}, C., {Baudin}, F., {et~al.} 2009, \nat, 459, 398

\bibitem[{{de Ridder} {et~al.}(2006){de Ridder}, {Barban}, {Carrier},
  {Mazumdar}, {Eggenberger}, {Aerts}, {Deruyter}, \&
  {Vanautgaerden}}]{deRidder06}
{de Ridder}, J., {Barban}, C., {Carrier}, F., {et~al.} 2006, \aap, 448, 689

\bibitem[{{Deubner}(1975)}]{Deubner75}
{Deubner}, F.~L. 1975, \aap, 44, 371

\bibitem[{{Dolginov} \& {Muslimov}(1984)}]{Dolginov84}
{Dolginov}, A.~Z. \& {Muslimov}, A.~G. 1984, \apss, 98, 15

\bibitem[{{Fletcher} {et~al.}(2006){Fletcher}, {Chaplin}, {Elsworth}, {Schou},
  \& {Buzasi}}]{Fletcher06}
{Fletcher}, S.~T., {Chaplin}, W.~J., {Elsworth}, Y., {Schou}, J., \& {Buzasi},
  D. 2006, \mnras, 824

\bibitem[{{Georgobiani} {et~al.}(2006){Georgobiani}, {Stein}, \&
  {Nordlund}}]{Georgobiani06}
{Georgobiani}, D., {Stein}, R.~F., \& {Nordlund}, {\AA}. 2006, in Astronomical
  Society of the Pacific Conference Series, Vol. 354, Solar MHD Theory and
  Observations: A High Spatial Resolution Perspective, ed. J.~{Leibacher},
  R.~F. {Stein}, \& H.~{Uitenbroek}, 109

\bibitem[{{Goldreich} \& {Keeley}(1977{\natexlab{a}})}]{Goldreich77a}
{Goldreich}, P. \& {Keeley}, D.~A. 1977{\natexlab{a}}, \apj, 211, 934

\bibitem[{{Goldreich} \& {Keeley}(1977{\natexlab{b}})}]{GK77}
{Goldreich}, P. \& {Keeley}, D.~A. 1977{\natexlab{b}}, \apj, 212, 243 (GK)

\bibitem[{{Goldreich} {et~al.}(1994){Goldreich}, {Murray}, \& {Kumar}}]{GMK94}
{Goldreich}, P., {Murray}, N., \& {Kumar}, P. 1994, \apj, 424, 466 (GMK)

\bibitem[{{Gough}(1976)}]{Gough76}
{Gough}, D. 1976, in Lecture notes in physics, Vol.~71, Problems of stellar
  convection, ed. E.~{Spiegel} \& J.-P. {Zahn} (Springer Verlag), 15

\bibitem[{{Gough}(1977)}]{Gough77}
{Gough}, D.~O. 1977, \apj, 214, 196

\bibitem[{{Gryanik} \& {Hartmann}(2002)}]{Gryanik02}
{Gryanik}, V. \& {Hartmann}, J. 2002, J. Atmos. Sci., 59, 2729

\bibitem[{{He} {et~al.}(2002){He}, {Rubinstein}, \& {Wang}}]{He02}
{He}, G.-W., {Rubinstein}, R., \& {Wang}, L.-P. 2002, Physics of Fluids, 14,
  2186

\bibitem[{{Houdek}(2006)}]{Houdek06}
{Houdek}, G. 2006, in ESA Special Publication, Vol. 624, Proceedings of SOHO
  18/GONG 2006/HELAS I, Beyond the spherical Sun, Published on CDROM, p. 28.1

\bibitem[{{Houdek}(2008)}]{Houdek08}
{Houdek}, G. 2008, Communications in Asteroseismology, 157, 137

\bibitem[{{Houdek} {et~al.}(1999){Houdek}, {Balmforth},
  {Christensen-Dalsgaard}, \& {Gough}}]{Houdek99}
{Houdek}, G., {Balmforth}, N.~J., {Christensen-Dalsgaard}, J., \& {Gough},
  D.~O. 1999, \aap, 351, 582

\bibitem[{{Jacoutot} {et~al.}(2008{\natexlab{a}}){Jacoutot}, {Kosovichev},
  {Wray}, \& {Mansour}}]{Jacoutot08b}
{Jacoutot}, L., {Kosovichev}, A.~G., {Wray}, A., \& {Mansour}, N.~N.
  2008{\natexlab{a}}, \apjl, 684, L51

\bibitem[{{Jacoutot} {et~al.}(2008{\natexlab{b}}){Jacoutot}, {Kosovichev},
  {Wray}, \& {Mansour}}]{Jacoutot08a}
{Jacoutot}, L., {Kosovichev}, A.~G., {Wray}, A.~A., \& {Mansour}, N.~N.
  2008{\natexlab{b}}, \apj, 682, 1386

\bibitem[{{Kjeldsen} \& {Bedding}(1995)}]{Kjeldsen95}
{Kjeldsen}, H. \& {Bedding}, T.~R. 1995, \aap, 293, 87

\bibitem[{{Kjeldsen} {et~al.}(2008){Kjeldsen}, {Bedding}, {Arentoft}, {Butler},
  {Dall}, {Karoff}, {Kiss}, {Tinney}, \& {Chaplin}}]{Kjeldsen08}
{Kjeldsen}, H., {Bedding}, T.~R., {Arentoft}, T., {et~al.} 2008, \apj, 682,
  1370

\bibitem[{{Kjeldsen} {et~al.}(2005){Kjeldsen}, {Bedding}, {Butler},
  {Christensen-Dalsgaard}, {Kiss}, {McCarthy}, {Marcy}, {Tinney}, \&
  {Wright}}]{Kjeldsen05}
{Kjeldsen}, H., {Bedding}, T.~R., {Butler}, R.~P., {et~al.} 2005, \apj, 635,
  1281

\bibitem[{{Kolmogorov}(1941)}]{Kolmogorov41}
{Kolmogorov}, A.~N. 1941, Dokl. Akad. Nauk SSSR, 30, 299

\bibitem[{{Kupka} \& {Robinson}(2007)}]{Kupka07}
{Kupka}, F. \& {Robinson}, F.~J. 2007, \mnras, 374, 305

\bibitem[{{Ledoux} \& {Walraven}(1958)}]{Ledoux58}
{Ledoux}, P. \& {Walraven}, T. 1958, in Handbuch der Physik, ed. F.~S., Vol.~51
  (Springer-Verlag (New York)), 353

\bibitem[{{Leibacher} \& {Stein}(1971)}]{Leibacher71}
{Leibacher}, J.~W. \& {Stein}, R.~F. 1971, \aplett, 7, 191

\bibitem[{Lesieur(1997)}]{Lesieur97}
Lesieur, M. 1997, Turbulence in fluids (Kluwer Academic Publishers)

\bibitem[{{Libbrecht}(1988)}]{Libbrecht88}
{Libbrecht}, K.~G. 1988, \apj, 334, 510

\bibitem[{{Lighthill}(1952)}]{Lighthill52}
{Lighthill}, M.~J. 1952, Proc. R. Soc. Lond., A211, 564

\bibitem[{{Michel} {et~al.}(2008){Michel}, {Baglin}, {Auvergne}, {Catala},
  {Samadi}, {Baudin}, {Appourchaux}, {Barban}, {Weiss}, {Berthomieu},
  {Boumier}, {Dupret}, {Garcia}, {Fridlund}, {Garrido}, {Goupil}, {Kjeldsen},
  {Lebreton}, {Mosser}, {Grotsch-Noels}, {Janot-Pacheco}, {Provost},
  {Roxburgh}, {Thoul}, {Toutain}, {Tiphene}, {Turck-Chieze}, {Vauclair},
  {Aerts}, {Alecian}, {Ballot}, {Charpinet}, {Hubert}, {Lignieres}, {Mathias},
  {Monteiro}, {Neiner}, \& {Poretti}}]{Michel08}
{Michel}, E., {Baglin}, A., {Auvergne}, M., {et~al.} 2008, Science, 322, 558

\bibitem[{{Mosser} {et~al.}(2009){Mosser}, {Michel}, {Appourchaux}, {Barban},
  {Baudin}, {Boumier}, {Bruntt}, {Catala}, {Deheuvels}, {Garc{\'{\i}}a},
  {Gaulme}, {Regulo}, {Roxburgh}, {Samadi}, {Verner}, {Auvergne}, {Baglin},
  {Ballot}, {Benomar}, \& {Mathur}}]{Mosser09}
{Mosser}, B., {Michel}, E., {Appourchaux}, T., {et~al.} 2009, \aap, 506, 33

\bibitem[{{Musielak} {et~al.}(1994){Musielak}, {Rosner}, {Stein}, \&
  {Ulmschneider}}]{Musielak94}
{Musielak}, Z.~E., {Rosner}, R., {Stein}, R.~F., \& {Ulmschneider}, P. 1994,
  \apj, 423, 474

\bibitem[{{Nordlund} \& {Stein}(1999)}]{Nordlund99b}
{Nordlund}, {\AA}. \& {Stein}, R.~F. 1999, in Astronomical Society of the
  Pacific Conference Series, Vol. 173, Stellar Structure: Theory and Test of
  Connective Energy Transport, ed. A.~{Gimenez}, E.~F. {Guinan}, \&
  B.~{Montesinos}, 91

\bibitem[{{Nordlund} \& {Stein}(2001)}]{Stein01I}
{Nordlund}, {\AA}. \& {Stein}, R.~F. 2001, \apj, 546, 576

\bibitem[{{Oboukhov}(1941)}]{Oboukhov41}
{Oboukhov}, A. 1941, Dokl. Akad. Sci. Nauk SSSR, 32, 22

\bibitem[{{Osaki}(1990)}]{Osaki90}
{Osaki}, Y. 1990, in Lecture Notes in Physics, Berlin Springer Verlag, Vol.
  367, Progress of Seismology of the Sun and Stars, ed. Y.~{Osaki} \&
  H.~{Shibahashi}, 75

\bibitem[{{Proctor} \& {Weiss}(1982)}]{Proctor82}
{Proctor}, M.~R.~E. \& {Weiss}, N.~O. 1982, Reports of Progress in Physics, 45,
  1317

\bibitem[{{Roca Cort{\'e}s} {et~al.}(1999){Roca Cort{\'e}s},
  {Monta{\~n}{\'e}s}, {Pall{\'e}}, {P{\'e}rez Hern{\'a}ndez}, {Jim{\'e}nez},
  {R{\'e}gulo}, \& {The GOLF Team}}]{Roca_Cortes99}
{Roca Cort{\'e}s}, T., {Monta{\~n}{\'e}s}, P., {Pall{\'e}}, P.~L., {et~al.}
  1999, in Astronomical Society of the Pacific Conference Series, Vol. 173,
  Stellar Structure: Theory and Test of Connective Energy Transport, ed.
  A.~{Gimenez}, E.~F. {Guinan}, \& B.~{Montesinos}, 305

\bibitem[{{Rosenthal} {et~al.}(1999){Rosenthal}, {Christensen-Dalsgaard},
  {Nordlund}, {Stein}, \& {Trampedach}}]{Rosenthal99}
{Rosenthal}, C.~S., {Christensen-Dalsgaard}, J., {Nordlund}, {\AA}., {Stein},
  R.~F., \& {Trampedach}, R. 1999, \aap, 351, 689

\bibitem[{{Samadi} {et~al.}(2009{\natexlab{a}}){Samadi}, {Belkacem}, {Goupil},
  {Dupret}, \& {Noels}}]{Samadi09c}
{Samadi}, R., {Belkacem}, K., {Goupil}, M.-J., {Dupret}, M.-A.and~{Brun}, A.,
  \& {Noels}, A. 2009{\natexlab{a}}, submitted to \apss

\bibitem[{{Samadi} {et~al.}(2008){Samadi}, {Belkacem}, {Goupil}, {Dupret}, \&
  {Kupka}}]{Samadi08}
{Samadi}, R., {Belkacem}, K., {Goupil}, M.~J., {Dupret}, M.-A., \& {Kupka}, F.
  2008, \aap, 489, 291

\bibitem[{{Samadi} {et~al.}(2007){Samadi}, {Georgobiani}, {Trampedach},
  {Goupil}, {Stein}, \& {Nordlund}}]{Samadi07a}
{Samadi}, R., {Georgobiani}, D., {Trampedach}, R., {et~al.} 2007, \aap, 463,
  297

\bibitem[{{Samadi} \& {Goupil}(2001)}]{Samadi00I}
{Samadi}, R. \& {Goupil}, M.~J. 2001, \aap, 370, 136 (SG)

\bibitem[{{Samadi} {et~al.}(2001){Samadi}, {Goupil}, \&
  {Lebreton}}]{Samadi00II}
{Samadi}, R., {Goupil}, M.~J., \& {Lebreton}, Y. 2001, \aap, 370, 147 

\bibitem[{{Samadi} {et~al.}(2006){Samadi}, {Kupka}, {Goupil}, {Lebreton}, \&
  {van't Veer-Menneret}}]{Samadi05b}
{Samadi}, R., {Kupka}, F., {Goupil}, M.~J., {Lebreton}, Y., \& {van't
  Veer-Menneret}, C. 2006, \aap, 445, 233

\bibitem[{{Samadi} {et~al.}(2009{\natexlab{b}}){Samadi}, {Ludwig}, {Belkacem},
  {Goupil}, {Benomar}, {Mosser}, {Dupret}, {Baudin}, {Appourchaux}, \&
  {Michel}}]{Samadi09b}
{Samadi}, R., {Ludwig}, H., {Belkacem}, K., {et~al.} 2009{\natexlab{b}}, \aap,
  in press (astro-ph/0910.4037)

\bibitem[{{Samadi} {et~al.}(2009{\natexlab{c}}){Samadi}, {Ludwig}, {Belkacem},
  {Goupil}, \& {Dupret}}]{Samadi09a}
{Samadi}, R., {Ludwig}, H., {Belkacem}, K., {Goupil}, M.-J., \& {Dupret}, M.
  2009{\natexlab{c}}, \aap, in press (astro-ph/0910.4027)


\bibitem[{{Samadi} {et~al.}(2003{\natexlab{a}}){Samadi}, {Nordlund}, {Stein},
  {Goupil}, \& {Roxburgh}}]{Samadi02II}
{Samadi}, R., {Nordlund}, {\AA}., {Stein}, R.~F., {Goupil}, M.~J., \&
  {Roxburgh}, I. 2003{\natexlab{a}}, \aap, 404, 1129

\bibitem[{{Samadi} {et~al.}(2003{\natexlab{b}}){Samadi}, {Nordlund}, {Stein},
  {Goupil}, \& {Roxburgh}}]{Samadi02I}
{Samadi}, R., {Nordlund}, {\AA}., {Stein}, R.~F., {Goupil}, M.~J., \&
  {Roxburgh}, I. 2003{\natexlab{b}}, \aap, 403, 303

\bibitem[{{Spiegel}(1962)}]{Spiegel62}
{Spiegel}, E. 1962, J. Geophys. Res., 67, 3063

\bibitem[{{Stein} {et~al.}(2004){Stein}, {Georgobiani}, {Trampedach}, {Ludwig},
  \& {Nordlund}}]{Stein04}
{Stein}, R., {Georgobiani}, D., {Trampedach}, R., {Ludwig}, H.-G., \&
  {Nordlund}, {\AA}. 2004, \solphys, 220, 229

\bibitem[{{Stein}(1967)}]{Stein67}
{Stein}, R.~F. 1967, Solar Physics, 2, 385

\bibitem[{{Stein} \& {Nordlund}(2001)}]{Stein01II}
{Stein}, R.~F. \& {Nordlund}, {\AA}. 2001, \apj, 546, 585

\bibitem[{{Ulrich}(1970)}]{Ulrich70}
{Ulrich}, R.~K. 1970, \apj, 162, 993

\bibitem[{{Unno} \& {Kato}(1962)}]{Unno62}
{Unno}, W. \& {Kato}, S. 1962, \pasj, 14, 417

\bibitem[{{Unno} {et~al.}(1989){Unno}, {Osaki}, {Ando}, {Saio}, \&
  {Shibahashi}}]{Unno89}
{Unno}, W., {Osaki}, Y., {Ando}, H., {Saio}, H., \& {Shibahashi}, H. 1989,
  Nonradial oscillations of stars (Tokyo: University of Tokyo Press, 1989, 2nd
  ed.)

\bibitem[{{V{\"o}gler} {et~al.}(2005){V{\"o}gler}, {Shelyag}, {Sch{\"u}ssler},
  {Cattaneo}, {Emonet}, \& {Linde}}]{Vogler05}
{V{\"o}gler}, A., {Shelyag}, S., {Sch{\"u}ssler}, M., {et~al.} 2005, \aap, 429,
  335

\bibitem[{{Xiong} {et~al.}(2000){Xiong}, {Cheng}, \& {Deng}}]{Xiong00}
{Xiong}, D.~R., {Cheng}, Q.~L., \& {Deng}, L. 2000, \mnras, 319, 1079

\end{thebibliography}

\end{document}